\documentclass{elsart}
\usepackage{epsfig}
\begin{document}
\begin{frontmatter}
\title{Nonleptonic $B$ decays into two light mesons in soft-collinear
  effective theory} 
\author{Junegone Chay\thanksref{email}},
\thanks[email]{E-mail address: chay@korea.ac.kr}
\author{Chul Kim}
\address{Department of Physics, Korea University, Seoul 136-701,
Korea} 
\begin{abstract}
We consider nonleptonic $B$ decays into two light mesons at leading
order in soft-collinear effective theory, and show that the
decay amplitudes  are factorized to all
orders in $\alpha_s$. The operators for
nonleptonic $B$ decays in the full theory are first matched to the 
operators in $\mathrm{SCET}_{\mathrm{I}}$, which is
the effective theory appropriate for $\sqrt{m_b
  \Lambda} <\mu <m_b$ with $\Lambda \sim 0.5$ GeV. We evolve the
operators and the relevant  time-ordered products in
$\mathrm{SCET}_{\mathrm{I}}$ to $\mathrm{SCET}_{\mathrm{II}}$, which is
appropriate for $\mu < \sqrt{m_b \Lambda}$. Using the 
gauge-invariant operators  in $\mathrm{SCET}_{\mathrm{II}}$, we
compute nonleptonic $B$ decays in SCET, including the
nonfactorizable spectator contributions and spectator
contributions to the heavy-to-light form factor. 
As an application, we present the decay
amplitudes for $\overline{B} \rightarrow \pi\pi$ in soft-collinear
effective theory.  
\end{abstract}

\end{frontmatter}

\section{Introduction}
Decay of $B$ mesons plays an important role in particle physics since
it is a testing ground for the Standard Model and a window for
possible new physics. We can obtain good information on the
Cabibbo-Kobayashi-Maskawa (CKM) matrix elements and CP
violation from $B$ decays. Since the contribution of QCD in $B$
decays changes the whole structure of the theory, the study of $B$
decays is an intertwined field of particle physics. Among these
decays, nonleptonic $B$ decays have been the subject of intense 
interest. Especially the treatment of nonperturbative effects from the
strong interaction is a serious theoretical problem in
nonleptonic decays. Precise experimental observation of 
nonleptonic $B$ decays makes it urgent to give  firm theoretical
prediction including the effects of CP violation. There has been a lot
of theoretical progress in nonleptonic $B$ decays and we suggest how
to consider nonleptonic $B$ decays from the viewpoint of the
soft-collinear effective theory (SCET).

The effective Hamiltonian for nonleptonic decays from the Standard
Model has been derived and the Wilson coefficients of the operators
for $B$ decays have been calculated to 
next-to-leading order, and next-to-next-to-leading order for some
operators \cite{buras}.  In order to evaluate the hadronic matrix
elements of four-quark operators, naive factorization \cite{naive}
was assumed, in which the matrix elements were reduced to products of
current matrix elements. But there was no justification for this
assumption except the argument of color transparency \cite{trans}.
Besides that, decay amplitudes depend on an arbitrary renormalization
scale $\mu$ in naive factorization since the Wilson coefficients
depend on $\mu$, while the matrix elements of operators do not.  
Ali et al. \cite{ali} have improved the problem of the scale
dependence by including radiative corrections of the operators before
taking hadronic matrix elements. Then the $\mu$ 
dependence of the Wilson coefficients is cancelled by that of the
radiative corrections of the operators. However, the decay
amplitudes depend on calculation schemes since the
off-shell renormalization scheme is used \cite{buras2}. 

Politzer and Wise \cite{politzer} suggested to take the heavy quark
mass limit in computing corrections to the decay rate ratio $\Gamma
(\overline{B} \rightarrow D^* \pi)/\Gamma( \overline{B} \rightarrow
D\pi)$. They have considered radiative corrections for nonfactorizable 
contributions and found that they are finite and the decay amplitudes
are factorized. Beneke et al. \cite{bbns} have extended this idea
to general two-body decays including two light final-state mesons, in
which the decay amplitudes can be expanded in a power series of
$1/m_b$. They show that nonfactorizable contributions including
spectator interactions are factorized as a convolution of the hard
scattering amplitudes with the  meson wave functions, and the
corrections are suppressed by powers of $1/m_b$. This is an important
step toward theoretical understanding of nonleptonic $B$ decays. First
the amplitudes can be obtained from first principles and a systematic
$1/m_b$ expansion is possible. Second, since the on-shell
renormalization is used, there is no scheme dependence. And they
improved previous approaches by including momentum-dependent parts,
which had not been included previously. However, when higher-twist
light-cone wave functions are included, there appear infrared
divergences in the amplitudes. These are treated as theoretical
uncertainties, but from the theoretical viewpoint it is a problem to
be solved in this approach.

Bauer et al. \cite{bauer1,bauer2} have proposed an effective field
theory in which massless quarks move with large energy. This
effective theory, called the soft-collinear effective theory, is
appropriate for light quarks with large energy. It has been applied to
hard scattering processes and $B$ decays
\cite{bauer5,bauer3,bauer4,chay1,manohar,bauer6}.    
It is also an appropriate effective theory for nonleptonic
$B$ decays to two light mesons. In this paper, we apply SCET to 
nonleptonic $B$ decays into light mesons. We construct all the relevant
operators in the effective theory at leading order in a
gauge-invariant way by integrating out all the off-shell modes. The
Wilson coefficients are calculated by matching the full theory onto
SCET. We show that the four-quark operators in SCET are factorized to
all orders in $\alpha_s$ and the argument of color transparency is 
explictly shown. And we consider nonfactorizable spectator
contribution in SCET, and find that they are 
also factorized to all orders in $\alpha_s$. 
The basic idea of the factorization properties in $B$ decays
into two light mesons was sketched in Ref.~\cite{chay2}. In this
paper, we extend the argument and discuss intricate characteristics of
SCET in nonleptonic decays, the details of the procedure of matching,
and present all the Wilson coefficients and technical details in the
calculation. We also present the analysis of $B\rightarrow \pi\pi$
decays as an application, which is consistent with the calculation in
the heavy quark mass limit at lowest order in $\alpha_s$. 

The organization of the paper is as follows:
In Section~\ref{sec2}, we consider the effective theories
$\mathrm{SCET}_{\mathrm{I}}$ and $\mathrm{SCET}_{\mathrm{II}}$, which
we employ in the analysis of nonleptonic $B$ decays and explain the
field contents and discuss how to match these effective theories. In
Section~\ref{sec3}, we construct four-quark operators in
$\mathrm{SCET}_{\mathrm{I}}$, which  
are gauge invariant by integrating out off-shell modes. This is
achieved by attaching gluons to fermion legs and by integrating out
off-shell intermediate states. We discuss the factorization properties
of the operators. In Section~\ref{sec4}, we match the full theory and
$\mathrm{SCET}_{\mathrm{I}}$ and obtain the Wilson coefficients of
the four-quark 
operators. In Section~\ref{sec5}, we consider nonfactorizable
spectator interaction, in which subleading operators are enhanced to
give the leading-order result. The structure of the subleading
operators and the mechanism for the enhancement are determined by the
power counting and the matching between effective theories.
In Section~\ref{sec6}, we consider the contributions to the
heavy-to-light form factor. The time-ordered products of  subleading
operators from the heavy-to-light current and the interaction of an
ultrasoft (usoft) quark and a collinear quark contribute to the form
factor. In Section~\ref{sec7} we combine
all the results to apply SCET to nonleptonic decays  $B\rightarrow \pi
\pi$. And finally a conclusion is presented. In Appendix A, we show
how the auxiliary field method can be used to derive gauge-invariant
four-quark operators in $\mathrm{SCET}_{\mathrm{II}}$. In Appendix B,
we employ the auxiliary field method to derive gauge-invariant
subleading operators in $\mathrm{SCET}_{\mathrm{I}}$.

\section{Construction of the effective theories
  $\boldmath\mathrm{SCET}_{\mathrm{I}}\unboldmath$ and 
  $\boldmath\mathrm{SCET}_{\mathrm{II}}\unboldmath$\label{sec2}}
The formalism of SCET and the procedure of the two-step matching are
discussed comprehensively in Refs.~\cite{bauer2,bauer6} and in
  Ref.~\cite{stewart}. We will not discuss them in detail, but we
review them briefly here.
The momentum of an energetic quark moving in the light-cone direction
$n^{\mu}$ can be decomposed as
\begin{equation}
p^{\mu} = \frac{n^{\mu}}{2} \overline{n} \cdot p + p_{\perp}^{\mu}
+\frac{\overline{n}^{\mu}}{2} n\cdot p= \mathcal{O} (\lambda^0)
+\mathcal{O} (\lambda) + \mathcal{O} (\lambda^2),
\end{equation}
where $n^{\mu}$, $\overline{n}^{\mu}$ are two light-like vectors
satisfying  $n^2=\overline{n}^2=0$, $n \cdot \overline n =2$. There
are three scales $\overline{n}\cdot p \sim Q$, $p_{\perp}^{\mu} \sim
Q\lambda$, and $n\cdot p \sim Q\lambda^2$, where $Q$ is the hard
scale. The momentum squared $p^2$ is typically of order
$Q\Lambda$, where $\Lambda$ is a typical hadronic scale $\Lambda
\sim$ 0.5 GeV. We introduce a small expansion parameter $\lambda \sim
\sqrt{\Lambda/Q}\ll 1$ to facilitate the power counting.
We construct SCET in two steps, namely $\mathrm{SCET}_\mathrm{I}$ for
$\sqrt{Q \Lambda} < \mu < Q$, and $\mathrm{SCET}_{\mathrm{II}}$
for $\mu <\sqrt{Q \Lambda}$. Formally we can construct
$\mathrm{SCET}_{\mathrm{II}}$ directly from the full theory, but it
is conceptually easy to construct $\mathrm{SCET}_\mathrm{I}$ and
$\mathrm{SCET}_{\mathrm{II}}$ successively. 

In SCET, the collinear quarks $\xi$  and
gluons $A_n^{\mu}$ have momenta $p^{\mu} =(n\cdot p, \overline{n}
\cdot p, p_{\perp}) \sim Q (\lambda^2, 1,\lambda)$. There are the soft
fields $q_s$, and $A_s^{\mu}$, with momenta $p_s^{\mu} \sim Q\lambda$,
and the usoft fields $q_{us}$, $A_{us}^{\mu}$ with momenta $p_{us}^{\mu} 
\sim Q\lambda^2$. The collinear quarks interact with collinear gluons or
usoft gluons, but not with soft gluons since they make
collinear particles or heavy quarks off the mass shell. In order to
derive the power counting 
method in $\mathrm{SCET}_{\mathrm{I}}$, we move all the 
dependence on $\lambda$ to the interaction vertices, and determine the
$\lambda$ dependence of all the fields \cite{bauer2}. Then we can
construct all the operators in powers of $\lambda$ systematically by
integrating out the off-shell modes of order $Q$. The guiding
principles of constructing operators in SCET are gauge invariance
\cite{bauer6}, reparameterization invariance \cite{chay1,manohar} and
the power counting \cite{bauer7}. The Wilson coefficients of the 
operators can be obtained through the matching between the full theory
and $\mathrm{SCET}_{\mathrm{I}}$  by requiring that the matrix 
elements of operators in both theories be the same at any order of the
perturbation theory. There is no mixing of operators with different
powers of $\lambda$ through radiative corrections as long as $Q\gg
\Lambda$ since the matching is performed perturbatively.

In $\mathrm{SCET}_{\mathrm{II}}$ for $\mu< \sqrt{Q\Lambda}$, we
integrate out all the off-shell modes of order
$\sqrt{Q\Lambda}$. Then the collinear fields in
$\mathrm{SCET}_{\mathrm{II}}$ have momenta $p^{\mu} \sim Q
(\lambda^{\prime 2}, 1, \lambda^{\prime})$, and the soft degrees of
freedom have momenta $p_s \sim Q\lambda^{\prime}$, where
$\lambda^{\prime} \sim \Lambda/Q$ is a new small expansion parameter
in $\mathrm{SCET}_{\mathrm{II}}$. Here 
the collinear modes have momenta $p^2 \sim \Lambda^2$, which is
appropriate to describe hadrons of size $\sim 1/\Lambda$. The degrees
of freedom of order $\sim \sqrt{Q\Lambda}$ in
$\mathrm{SCET}_{\mathrm{I}}$ are all integrated out, and the usoft
degrees of freedom in $\mathrm{SCET}_{\mathrm{I}}$ of order $p_{us}
\sim Q\lambda^2$ remain in $\mathrm{SCET}_{\mathrm{II}}$ with momentum
$p_s \sim Q\lambda^2 = Q\lambda^{\prime}$, and we call them soft
degrees of freedom in $\mathrm{SCET}_{\mathrm{II}}$. The matching of
$\mathrm{SCET}_{\mathrm{I}}$ onto $\mathrm{SCET}_{\mathrm{II}}$ is
performed at $\mu \sim \sqrt{Q\Lambda}$. 

A subtle technical point in matching is to require gauge
invariance. In order to see how to keep the gauge invariance, let us
consider a simple example of a heavy-to-light current. In the full
theory the current $\overline{q} \Gamma b$ is matched to the operator
in $\mathrm{SCET}_{\mathrm{I}}$ as
\begin{equation}
C(n\cdot \mathcal{P}) \overline{\xi} \Gamma W h,
\end{equation}
where $C(n\cdot \mathcal{P})$ is the Wilson coefficient, and
$\mathcal{P}^{\mu} = \half n^{\mu} n\cdot \mathcal{P}
+\mathcal{P}_{\perp}^{\mu}$ is the label momentum operator. The factor
$W$ is the Wilson line which is given by
\begin{equation}
W= \sum_{\mathrm{perm}} \exp \Bigl[ -g \frac{1}{\overline{n} \cdot
    \mathcal{P}} \overline{n}\cdot A_n \Bigr].
\end{equation}
This Wilson line is obtained by attaching collinear gluons to the
heavy quark and integrate out the off-shell modes of the intermediate
states of order $Q$. 

In $\mathrm{SCET}_{\mathrm{II}}$, the emission of soft gluons $p^{\mu}
\sim Q\lambda^{\prime}$ makes the collinear fields off the mass shell
and we have to integrate out the off-shell modes. This is achieved by
attaching usoft gluons to external fermion lines in
$\mathrm{SCET}_{\mathrm{I}}$ and by integrating out the off-shell
modes of order $Q\lambda^2 =Q\lambda^{\prime}$. It corresponds to
factorizing the usoft-collinear interactions with the field
redefinition \cite{bauer6}  
\begin{equation}
\xi^{(0)} = Y^{\dagger} \xi, \ \ A_n^{(0)} = Y^{\dagger}A_n Y,
\ \ 
Y(x) = \mathrm{P} \exp \Bigl( ig \int_{-\infty}^x ds n\cdot
  A_{us} (ns) \Bigr).
\label{redef1}
\end{equation}
The collinear effective Lagrangian can be written in terms of
$\xi^{(0)}$ and $A_n^{(0)\mu}$, that is, the collinear-usoft
interactions are factorized with the field redefinitions given in
Eq.~(\ref{redef1}). Then we match $\mathrm{SCET}_{\mathrm{I}}$ onto
$\mathrm{SCET}_{\mathrm{II}}$ at a scale $\mu\sim
\sqrt{Q\Lambda}$. For example, the heavy-to-light current is matched
as
\begin{equation}
\overline{q} \Gamma b \rightarrow C(\overline{n} \cdot \mathcal{P})
\overline{\xi} W \Gamma h \rightarrow  C(\overline{n} \cdot
\mathcal{P}) \overline{\xi}^{(0)} W^{(0)}  \Gamma Y^{\dagger} h,
\end{equation}
where $W^{(0)} = Y^{\dagger} W Y$. In $\mathrm{SCET}_{\mathrm{II}}$,
the fields with the superscript $(0)$ are fundamental objects.
We rename the usoft field as the soft field in
$\mathrm{SCET}_{\mathrm{II}}$. We write $Y^{\dagger} h 
\rightarrow S^{\dagger} h$, and lower the off-shellness of the
collinear fields.

Since the leading collinear Lagrangian is the same in
$\mathrm{SCET}_{\mathrm{I}}$ and $\mathrm{SCET}_{\mathrm{II}}$, we can
simply replace $C(\overline{n}\cdot \mathcal{P}) \overline{\xi}^{(0)}
W^{(0)} \rightarrow C(\overline{n}\cdot \mathcal{P})
\overline{\xi}^{\mathrm{II}} W^{\mathrm{II}}$, and we obtain the final
form of the operator 
\begin{equation}
\overline{q} \Gamma b \rightarrow C(\overline{n} \cdot \mathcal{P})
\overline{\xi}^{\mathrm{II}} W^{\mathrm{II}} \Gamma S^{\dagger} h,
\end{equation}
where the superscript $\mathrm{II}$, which denotes
$\mathrm{SCET}_{\mathrm{II}}$, will be dropped from now on.
From the two-step matching, it is clear why the Wilson coefficients do
not depend on the soft momentum. 

The advantage of the two-step matching is manifest when
we consider time-ordered products in $\mathrm{SCET}_{\mathrm{I}}$
since these can induce jet functions involving the fluctuations of
order $p^2\sim Q\Lambda$. In $\mathrm{SCET}_{\mathrm{I}}$ we can
compute the jet functions with a well-defined set of Feynman rules
independent of the computation of the Wilson coefficients at $p^2
=Q^2$. And the scaling of operators in $\mathrm{SCET}_{\mathrm{II}}$
is constrained by the power counting in $\mathrm{SCET}_{\mathrm{I}}$
and especially,  $\mathrm{SCET}_{\mathrm{I}}$ puts a constraint of
the number of factors of $1/\Lambda$, which can be induced from the
fluctuations $1/Q\Lambda$. Therefore we can know
how many powers of $1/Q\Lambda$ appear for a given process at a given
order in $\alpha_s$, and this power counting is not spoiled by the
loop effects since there is no dependence on the soft momentum in the
Wilson coefficients. This feature will be seen explicitly when
we evaluate the decay amplitudes of nonleptonic decays which involve
time-ordered products.

In nonleptonic decays into two light mesons, there are collinear
quarks moving in opposite directions. We choose these two directions
as $n^{\mu}$ and $\overline{n}^{\mu}$. For an energetic quark moving
in the $\overline{n}^{\mu}$ direction, the momentum is decomposed as
\begin{equation}
p_{\overline{n}}^{\mu} = \frac{n \cdot p_{\overline{n}}}{2}
    \overline{n}^{\mu} +     p_{\overline{n}\perp}^{\mu} 
+\frac{\overline{n}\cdot p_{\overline{n}}}{2} n^{\mu} = \mathcal{O}
    (\lambda^0) + \mathcal{O} (\lambda) + \mathcal{O} (\lambda^2).
\end{equation}
We denote a collinear quark by $\xi$ ($\chi$) which moves in
the $n^{\mu}$ ($\overline{n}^{\mu}$) direction. These fields satisfy
the relations
\begin{equation} 
\FMslash{n} \xi = 0, \ \ \frac{\FMslash{n}\FMslash{\overline{n}}}{4}
  \xi   = \xi,  \ \ 
\FMslash{\overline{n}} \chi = 0, \ \  \frac{\FMslash{\overline{n}}
  \FMslash{n}}{4} \chi = \chi.
\end{equation}
The collinear gauge field in the $n^{\mu}$ ($\overline{n}^{\mu}$)
direction is written as $A_n^{\mu}$ ($A_{\overline{n}}^{\mu}$).
The effective Lagrangian for $\chi$ and
$A_{\overline{n}}^{\mu}$, which can be obtained from the colllinear
Lagrangian for $\xi$ and $A_n^{\mu}$ by replacing $\xi 
\leftrightarrow \chi$, $n^{\mu}\leftrightarrow \overline{n}^{\mu}$
respectively. This situation is simlar to two jets in the opposite
direction \cite{bauer4}, but we have a heavy quark  
interacting with collinear gluons in both directions, which makes the
analysis more interesting and complicated.

\section{Construction of gauge-invariant four-quark
  operators in SCET\label{sec3}} 

The effective Hamiltonian for $B$ decays in the full theory is given
as 
\begin{equation}
H_{\mathrm{eff}} = \frac{G_F}{\sqrt{2}} \sum_{p=u,c} V_{pd}^* V_{pb}
\Bigl( C_1 O_1^p + C_2 O_2^p +\sum_{i=3,\cdots,6,8} C_i O_i\Bigr),
\label{fullop}
\end{equation}
where the local $\Delta B=1$ operators are given by
\begin{eqnarray}
O_1^p &=& (\overline{p}_{\alpha} b_{\alpha})_{V-A}
(\overline{d}_{\beta} p_{\beta})_{V-A}, \ \ O_2^p =
(\overline{p}_{\beta} b_{\alpha})_{V-A} 
(\overline{d}_{\alpha} p_{\beta})_{V-A}, \nonumber \\
O_3 &=& (\overline{d}_{\alpha} b_{\alpha})_{V-A} \sum_q
(\overline{q}_{\beta} q_{\beta})_{V-A},  \ \ O_4 =
(\overline{d}_{\beta} b_{\alpha})_{V-A} \sum_q 
(\overline{q}_{\alpha} q_{\beta})_{V-A}, \nonumber \\
O_5 &=& (\overline{d}_{\alpha} b_{\alpha})_{V-A} \sum_q
(\overline{q}_{\beta} q_{\beta})_{V+A},  \ \ O_6 =
(\overline{d}_{\beta} b_{\alpha})_{V-A} \sum_q 
(\overline{q}_{\alpha} q_{\beta})_{V+A}, \nonumber \\
O_8 &=& \frac{-g}{8\pi^2} m_b \overline{d}_{\alpha} \sigma^{\mu\nu}
(1+\gamma_5) (T_a)_{\alpha\beta} b_{\beta} G_{\mu\nu}^a.
\end{eqnarray}
Here $p=u,c$ and $d$ denotes down-type quarks, $G_{\mu\nu}^a$ is
the chromomagnetic field strength tensor, and $T_a$ are the color
$SU(3)$ generators.

The process of obtaining the gauge-invariant operators
in SCET requires two-step matching \cite{neuts,form} because the SCET
involves two different scales $\mu\sim m_b$ and $\mu_0\sim 
  \sqrt{m_b  \Lambda}$. First we match the 
full theory  onto $\mathrm{SCET}_{\mathrm{I}}$ at $\mu = m_b$, and
we match successively onto $\mathrm{SCET}_{\mathrm{II}}$ at $\mu_0$. A
concrete example of the two-step matching was illustrated in
Ref.~\cite{pirjol}.  
In order to construct the operators in SCET, we first have to specify
which quark or antiquark goes to a certain direction to form a light
meson. We set $\overline{n}^{\mu}$ as the direction of a
quark-antiquark pair to form a light meson, and $n^{\mu}$ as the
direction of the remaining quark which combines with a spectator quark
in a $B$ meson to form another meson. Therefore the construction of
the operators is process-dependent in the sense that we first specify
the direction of each outgoing quark, and the number of operators in
SCET is doubled because we have two possibilities to assign two quark
fields in both directions. 

A generic four-quark operator for nonleptonic $B$ decays in SCET has
the form $(\overline{\xi} \Gamma_1 h) (\overline{\chi} \Gamma_2
\chi)$, or $(\overline{\chi} \Gamma_1 h)(\overline{\xi} \Gamma_2
\chi)$, where  $\Gamma_1$ and $\Gamma_2$ are Dirac matrices,
and $h$ is the heavy quark field in the heavy quark effective
theory. These operators are derived from the operator $\overline{q}_1 
\Gamma_1 b \cdot \overline{q}_2 \Gamma_2 q_3$ in the full theory where
$q_i$ ($i=1,2,3$) are light quarks. The operator $(\overline{\xi}
\Gamma_1 h) (\overline{\chi} \Gamma_2 \chi)$ is obtained by
replacing $q_1$ by $\xi$, and $q_2$ and $q_3$ by $\chi$. For the
operator $(\overline{\chi} \Gamma_1 h) (\overline{\xi} \Gamma_2
\chi)$, we replace $q_2$ by $\xi$ and $q_1$ and $q_3$ by the $\chi$
fields. The second operator produces a light meson, in which one quark
comes from the heavy-to-light current and another antiquark from
the light-to-light current to form a meson in the $\overline{n}^{\mu}$
direction. A remaining quark goes to the $n^{\mu}$ direction. We can
transform this operator to the form $(\overline{\xi} \Gamma_1^{\prime}
h) (\overline{\chi} \Gamma_2^{\prime} \chi)$ by Fierz
transformation. In order to simplify the organization of the
computation, all the operators will be written in the form
$(\overline{\xi} \Gamma_1  h)(\overline{\chi}\Gamma_2 \chi)$. If a
Fierz transformation is necessary to obtain this form, we apply Fierz
transformation in the full theory and perform matching
accordingly. Therefore we only have to consider operators of the form
$(\overline{\xi} \Gamma_1 h)(\overline{\chi} \Gamma_2  \chi)$, but the
two types of the operators with or without Fierz transformation are
regarded as distinct and their Wilson coefficients are
different.

Though we know the form of the operators, the operator
$(\overline{\xi} \Gamma_1 h)(\overline{\chi} \Gamma_2 \chi)$ itself
is not gauge invariant under the collinear, soft and usoft gauge  
transformations. In order to obtain gauge-invariant operators in
$\mathrm{SCET}_{\mathrm{II}}$, we employ two-step matching.
We consider the emissions of collinear gluons $A_n^{\mu}$ and
$A_{\overline{n}}^{\mu}$ from external fermions in the full theory and
integrate out the off-shell intermediate states of
order $\sim m_b$ to obtain the collinear gauge-invariant operators in
$\mathrm{SCET}_{\mathrm{I}}$. Then we go down to
$\mathrm{SCET}_{\mathrm{II}}$, in which we consider the emissions of
soft gluons $A_s^{\mu}$ which have momenta of order $\sim m_b
\lambda^{\prime} = m_b \lambda^2$. The emission of soft gluons can
cause intermediate states off the mass shell by the amount $p^2 \sim
m_b \Lambda$, which should be integrated out to obtain the
final form of the operators in $\mathrm{SCET}_{\mathrm{II}}$. To
simplify the calculation, we will directly match the full-QCD
operators to the operators in $\mathrm{SCET}_{\mathrm{II}}$ by
considering the emissions of the collinear and the soft gluons from
each fermion in the full QCD and integrate out all the intermediate
states of order $\sim m_b$ and $\sqrt{m_b \Lambda}$ to obtain the
gauge-invariant operators in $\mathrm{SCET}_{\mathrm{II}}$, as was
done in Ref.~\cite{bauer6}. The result is equivalent to the two-step
matching described above. We will use the two-step matching explicitly
in treating the time-ordered products.

\begin{figure}[t]
\begin{center}
\epsfig{file=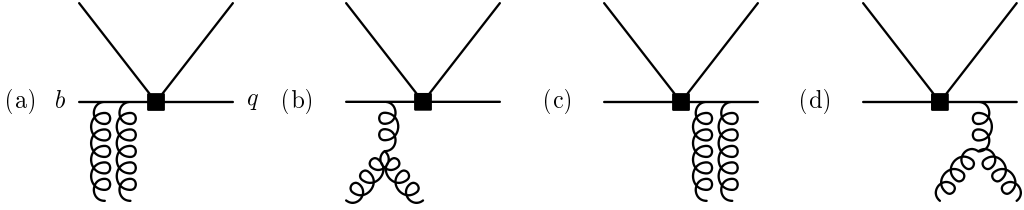, width=13.5cm}
\end{center}
\caption{QCD diagrams attaching two gluons to external fermions to
integrate out off-shell modes. The external gluons $A_n$,
$A_{\overline{n}}$ or $A_s$  make the intermediate
states off the mass shell.  Diagrams with gluons attached to other
fermions are omitted.}  
\label{fig1}
\end{figure}

We can construct gauge-invariant operators by
attaching collinear or soft gluons to each fermion in the full QCD
and integrate out off-shell modes. Note that the soft gluons here have
momenta of order $\sim m_b \lambda^{\prime} = m_b \lambda^2$, defined
in $\mathrm{SCET}_{\mathrm{II}}$. Because the ordering of gauge fields
is important due to the nonabelian nature of the gauge fields, we
consider corrections to order $g^2$. Typical Feynman
diagrams at order $g^2$ are shown in Fig.~\ref{fig1}, and the remaining
diagrams in which gluons are attached to other fermions
are omitted. But it is straightforward to attach two gluons 
to all the fermion lines making intermediate states off-shell. 

If collinear and soft gluons are attached to the heavy quark
[Fig.~\ref{fig1} (a), (b)], the intermediate heavy quark and gluon states
are off the mass shell and we integrate them out. For the collinear
quark $\xi$, the interaction with $A_{\overline{n}}^{\mu}$or
$A_s^{\mu}$ makes the collinear quark off the mass shell. And
the interaction with $A_n^{\mu}$ or $A_s^{\mu}$ makes the $\chi$
field off the mass shell. We add all the possible configurations 
in which the intermediate states become off the mass shell
and collect the terms at leading order in $\Lambda$.

Let us introduce the factors
\begin{equation}
A=\frac{\overline{n} \cdot A_n}{\overline{n}\cdot q_n}, \ B=
\frac{n\cdot A_{\overline{n}}}{n\cdot q_{\overline{n}}}, \ C=
\frac{\overline{n} \cdot A_s}{\overline{n} \cdot q_s}, \ D=
\frac{n\cdot A_s}{n\cdot q_s},
\end{equation}
where $q_n^{\mu}$ ($q_{\overline{n}}^{\mu}$) is the momentum of the
collinear gluon $A_n^{\mu}$ ($A_{\overline{n}}^{\mu}$), and
$q_s^{\mu}$ is the soft momentum of the soft gluon $A_s^{\mu}$. 
The Wilson lines corresponding to each type of gluons are obtained by
exponentiating the above factors as
\begin{eqnarray}
W&=& \sum_{\mathrm{perm}} \exp \Bigl[ -g \frac{1}{\overline{n} \cdot
    \mathcal{P}} \overline{n}\cdot A_n \Bigr], \ \ \overline{W}=
    \sum_{\mathrm{perm}}  \exp \Bigl[ -g \frac{1}{n \cdot
    \mathcal{Q}} n\cdot A_{\overline{n}} \Bigr], \nonumber \\
\overline{S}&=&
    \sum_{\mathrm{perm}}  \exp \Bigl[ -g \frac{1}{\overline{n} \cdot
    \mathcal{R}} \overline{n}\cdot A_s \Bigr], \ \ 
S= \sum_{\mathrm{perm}} \exp \Bigl[ -g \frac{1}{n \cdot
    \mathcal{R}}n \cdot A_s \Bigr]. 
\end{eqnarray}
Here $\mathcal{P}^{\mu} = \overline{n}\cdot \mathcal{P} n^{\mu}/2 +
\mathcal{P}_{\perp}^{\mu}$ ($\mathcal{Q}^{\mu} = 
n\cdot \mathcal{Q} \overline{n}^{\mu}/2 + \mathcal{Q}_{\perp}^{\mu}$)
is the label momentum operator for 
collinear fields in the $n^{\mu}$ ($\overline{n}^{\mu}$) direction, 
and the operator $\mathcal{R}$ is the
operator extracting the soft momentum from soft fields.

When we sum over all these diagrams, the color singlet operators
of the form $(\overline{\xi}_{\alpha} \Gamma_1 h_{\alpha})
(\overline{\chi}_{\beta} \Gamma_2 \chi_{\beta})$ and the nonsinglet
operators of the form  $(\overline{\xi}_{\beta} \Gamma_1 h_{\alpha})
(\overline{\chi}_{\alpha} \Gamma_2 \chi_{\beta})$ are affected by the
gauge fields differently and the  final form of the four-quark
operators can be written as 
\begin{eqnarray}
O_S&=& (\overline{\xi}_{\alpha} \Gamma_1
h_{\alpha})(\overline{\chi}_{\beta} \Gamma_2 \chi_{\beta}) \rightarrow
H_{\alpha\beta}^S L_{\gamma\delta}^S  (\overline{\xi}_{\alpha} \Gamma_1
h_{\beta})(\overline{\chi}_{\gamma} \Gamma_2 \chi_{\delta}), \nonumber
\\
O_N&=& (\overline{\xi}_{\beta} \Gamma_1
h_{\alpha})(\overline{\chi}_{\alpha} \Gamma_2 \chi_{\beta}) \rightarrow
H_{\gamma\beta}^N L_{\alpha\delta}^N  (\overline{\xi}_{\alpha} \Gamma_1
h_{\beta})(\overline{\chi}_{\gamma} \Gamma_2 \chi_{\delta}),
\end{eqnarray}
where $H_{\alpha \beta}^O$, $L_{\gamma\delta}^O$ ($O=S,N$) are the color
factors.

One interesting case arises when we attach $A_n^{\mu}$ and
$A_{\overline{n}}^{\mu}$ to a heavy quark. At leading order in $\Lambda$,
the  amplitude in Fig.~\ref{fig1} (a), with $A_n^{\mu}$ and
$A_{\overline{n}}^{\mu}$ carrying the incoming momentum $q_n^{\mu}$ and
$q_{\overline{n}}^{\mu}$ respectively, is given by  
\begin{eqnarray}
M_a&=& \frac{g^2}{m_b (\overline{n} \cdot q_n +n \cdot q_{\overline{n}}
  ) +\overline{n}\cdot q_n n\cdot q_{\overline{n}}} \nonumber \\
&&\times \overline{q}
  \Gamma_1 \Bigl[ \Bigl( m_b n\cdot A_{\overline{n}} + \overline{n}
  \cdot q_n n\cdot A_{\overline{n}} \frac{\FMslash{n}
  \FMslash{\overline{n}}}{4} \Bigr) \frac{\overline{n} \cdot
  A_n}{\overline{n} \cdot q_n} \nonumber \\
&&+\Bigl(m_b \overline{n} \cdot A_n + n\cdot q_{\overline{n}}
  \overline{n} \cdot A_n \frac{\FMslash{\overline{n}} \FMslash{n}}{4}
  \Bigr) \frac{n\cdot A_{\overline{n}}}{n\cdot q_{\overline{n}}}
  \Bigr] b,
\end{eqnarray}
and the amplitude for Fig.~\ref{fig1} (b) with $A_n^{\mu}$ and
$A_{\overline{n}}^{\mu}$ is written as
\begin{eqnarray}
M_b &=& \frac{ig^2}{2} f_{abc} \overline{q} \Gamma_1 T_a \frac{n\cdot
  A_{\overline{n}}^b \overline{n} \cdot A_n^c}{n\cdot q_{\overline{n}}
  \overline{n} \cdot q_n}b 
-\frac{ig^2 f_{abc}}{m_b (\overline{n} \cdot q_n +n \cdot
  q_{\overline{n}} ) +\overline{n}\cdot q_n n\cdot q_{\overline{n}}}
  \nonumber \\ 
&&\times \overline{q} \Gamma_1 T^a n\cdot A_{\overline{n}}^b
  \frac{\overline{n} \cdot A_n^c}{\overline{n} \cdot q_n} \Bigl( m_b
  +\overline{n} \cdot q_n \frac{\FMslash{n} \FMslash{\overline{n}}}{4}
  \Bigr) b.
\end{eqnarray}
At first sight, these amplitudes contain  complicated denominators
which cannot be expressed in terms of $A$, $B$, $C$ or $D$. However,
if we add $M_a$ and $M_b$, we obtain
\begin{equation}
M_a + M_b =\frac{ig^2}{2} f_{abc} \overline{q} \Gamma_1 T_a \frac{n\cdot
  A_{\overline{n}}^b \overline{n} \cdot A_n^c}{n\cdot q_{\overline{n}}
  \overline{n} \cdot q_n}b+ g^2 \overline{q} \Gamma_1
  \frac{\overline{n} \cdot A_n}{\overline{n} 
  \cdot q_n} \frac{n\cdot A_{\overline{n}}}{n\cdot q_{\overline{n}}}
  b,
\label{heavy}
\end{equation}
where there appear only $A$ and
$B$. Therefore the role of the triple gluon vertex is critical not
only in determining the order of the Wilson lines but also in making
the final expression simple. We will derive gauge-invariant operators
using the auxiliary fields in Appendix A, and this cancellation is
important in constructing the Lagrangian for the auxiliary fields. 

For singlet operators, when we sum over all the Feynman diagrams, we
have  
\begin{equation}
H^S_{\alpha\beta} = \Bigl[ 
 g(-A+D) -g^2 AD \Bigr]_{\alpha \beta}, \ L^S_{\gamma\delta}
 =\delta_{\gamma\delta}, 
\label{dfsing}
\end{equation}
to order $g^2$. Here we show only the products of two different
gauge fields, since they indicate the ordering
of the Wilson lines. From Eq.~(\ref{dfsing}), $H^S_{\alpha\beta}$
suggests the Wilson lines to be $(WS^{\dagger})_{\alpha\beta}$. 
For color nonsiglet operators, we have
\begin{eqnarray}
H^N_{\gamma\beta} &=&\Bigl[ g(-B+C) -g^2 BC\Bigr]_{\gamma\beta},
\nonumber \\ 
L^N_{\alpha\delta} &=& \Bigl[g(-A+B-C+D) \nonumber \\
&&+g^2 (AC+DB-CB-DC-AB-AD)\Bigr]_{\alpha\delta}, 
\label{dfnon}
\end{eqnarray}
where $H^N_{\gamma\beta}$ suggests the Wilson line in
the order $(\overline{W} \overline{S}^{\dagger})_{\gamma\beta}$, and
$L^N_{\alpha\delta}$ suggests the form $(WS^{\dagger}\overline{S}
\overline{W}^{\dagger})_{\alpha\delta}$. Therefore the
gauge-invariant singlet and nonsinglet operators in
$\mathrm{SCET}_{\mathrm{II}}$ are given by
\begin{eqnarray}
O_S &=& \Bigl[(\overline{\xi} W)_{\alpha} \Gamma_1
(S^{\dagger}h)_{\alpha} \Bigr]\cdot 
\Bigl[(\overline{\chi}\overline{W})_{\beta} \Gamma_2
(\overline{W}^{\dagger} \chi)_{\beta}\Bigr], \nonumber \\
O_N &=& \Bigl[
(\overline{\xi} WS^{\dagger} \overline{S})_{\beta} \Gamma_1
(\overline{S}^{\dagger} h )_{\alpha} \Bigr]\cdot
\Bigl[ (\overline{\chi} \overline{W})_{\alpha} \Gamma_2
(\overline{W}^{\dagger} \chi)_{\beta} \Bigr].
\label{gio2}
\end{eqnarray}
The operator $O_N$ can be written as $\Bigl(
(\overline{\xi} WS^{\dagger})_{\beta} \Gamma_1
(\overline{S}^{\dagger} h )_{\alpha} \Bigr)
\Bigl( (\overline{\chi} \overline{W})_{\alpha} \Gamma_2
(\overline{S}\overline{W}^{\dagger} \chi)_{\beta} \Bigr)$, but this
is identical to $O_N$ since $(WS^{\dagger}
\overline{S})_{\alpha\gamma} \otimes
(\overline{W}^{\dagger})_{\gamma\beta} =
(WS^{\dagger})_{\alpha\gamma}  \otimes
(\overline{S}\overline{W}^{\dagger})_{\gamma\beta}$.

All the four-quark operators for $B$ decays are of the
form $O_S$ or $O_N$ with different Dirac structure. And the form of
the operators $O_S$ and $O_N$ in Eq.~(\ref{gio2}) manifestly shows the
factorization of four-quark operators at leading order in SCET and to
all orders in $\alpha_s$. In the operators $O_S$ and $O_N$, the
interactions of $A_n^{\mu}$ and $A_s^{\mu}$ occur only in the
heavy-to-light current sector, while the interactions of 
$A_{\overline{n}}^{\mu}$ occur only in the light-to-light current
sector. If a collinear or a soft gluon is emitted from one sector and
is absorbed by the other sector, the interaction vanishes, or if it
does not vanish, the momentum transfer is of order $\sqrt{m_b
  \Lambda}$, which is already integrated out to produce the 
Wilson coefficients or jet functions. That is, the gluon exchange
between the two current sectors is not allowed, and the gluon exchange
is allowed only in each sector.  Due to 
this property, the form of the operators is preserved even though
there are any possible exchange of gluons to all orders. The
Wilson coefficients may be different at different orders of the
perturbation theory. This is an
explicit proof of color transparency in SCET, and we can safely
calculate the matrix elements of the operators in terms of a product
of the matrix elements of the two currents.  

Note that the terminology ``factorization'' is used in two
ways. First, it means that the matrix elements of the
four-quark operators reduce to products of the matrix elements of
the currents, and the matrix element can be written as
\begin{equation}
\langle M_1 M_2 | j_1 \otimes j_2 |B\rangle = \langle M_1 |j_1 |B\rangle
\langle M_2|j_2 |0\rangle.
\label{facto}
\end{equation}
This was first assumed in naive factorization. It is possible only
when there is no gluon exchange which connects the two currents and it
is explicitly shown in SCET to all orders in $\alpha_s$.  

Another meaning of factorization appears in high-energy processes in
which a physical amplitude can be separated into a short-distance
part and a long-distance part. For example, exclusive hadronic form
factors at momentum transfer $Q^2 \gg \Lambda^2$ factor
into nonperturbative light-cone wave functions $\phi$ for mesons,
convoluted with a hard scattering kernel $T$ as \cite{brodsky}
\begin{equation}
F(Q^2) =\frac{f_a f_b}{Q^2} \int dx dy T(x,y,\mu) \phi_a (x,\mu)
\phi_b (y,\mu) +\cdots.
\end{equation}
We discuss both types of factorization in this paper. So far, we have
considered the factorization of matrix elements. The second meaning of
factorization will be discussed when we consider spectator
contributions.

By matching the four-quark operators in the full theory onto SCET, the 
effective Hamiltonian for $B$ decays in SCET can be written as 
\begin{eqnarray}
H_{\mathrm{eff}} &=& \frac{G_F}{\sqrt{2}} \sum_{T=R,C}
\Bigl[ V_{ub} V_{ud}^* \Bigl(
  C_{1T} O_{1T} +   C_{2T} O_{2T} \Bigr)\nonumber \\
&&+ \sum_{p=u,c} V_{pb} V_{pd}^*  \sum_{i=3,\cdots,6} C_{iT}^p 
  O_{iT} \Bigr) \Bigr], 
\label{scetham}
\end{eqnarray}
where
\begin{eqnarray}
O_{1R} &=& \Bigl[ (\overline{\xi}^u W)_{\alpha} (S^{\dagger}
h)_{\alpha} \Bigr]_{V-A} \Bigl[ (\overline{\chi}^d \overline{W})_{\beta}
(\overline{W}^{\dagger}\chi^u)_{\beta} \Bigr]_{V-A}, \nonumber \\
O_{2R} &=&  \Bigl[ (\overline{\xi}^u
W S^{\dagger} \overline{S})_{\beta} (\overline{S}^{\dagger}
h)_{\alpha} \Bigr]_{V-A} \Bigl[ (\overline{\chi}^d \overline{W}
  )_{\alpha} (\overline{W}^{\dagger}\chi^u)_{\beta} \Bigr]_{V-A},
\nonumber \\ 
O_{3R} &=& \Bigl[ (\overline{\xi}^d W)_{\alpha} (S^{\dagger}
h)_{\alpha} \Bigr]_{V-A} \sum_q \Bigl[ (\overline{\chi}^q
\overline{W})_{\beta} (\overline{W}^{\dagger}\chi^q)_{\beta}
\Bigr]_{V-A}, \nonumber \\ 
O_{4R} &=&  \Bigl[ (\overline{\xi}^d
W S^{\dagger} \overline{S})_{\beta} (\overline{S}^{\dagger}
h)_{\alpha} \Bigr]_{V-A} \sum_q \Bigl[ (\overline{\chi}^q \overline{W}
)_{\alpha} 
(\overline{W}^{\dagger}\chi^q)_{\beta} \Bigr]_{V-A}, \nonumber \\
O_{5R} &=& \Bigl[ (\overline{\xi}^d W)_{\alpha} (S^{\dagger}
h)_{\alpha} \Bigr]_{V-A} \sum_q \Bigl[ (\overline{\chi}^q
\overline{W})_{\beta} (\overline{W}^{\dagger} \chi^q)_{\beta}
\Bigr]_{V+A}, \nonumber \\ 
O_{6R} &=&  \Bigl[ (\overline{\xi}^d
W S^{\dagger} \overline{S})_{\beta} (\overline{S}^{\dagger}
h)_{\alpha} \Bigr)]_{V-A} \sum_q \Bigl[ (\overline{\chi}^q \overline{W}
)_{\alpha} 
(\overline{W}^{\dagger}\chi^q)_{\beta} \Bigr]_{V+A}, \nonumber \\
O_{1C} &=&  \Bigl[ (\overline{\xi}^d
W S^{\dagger} \overline{S})_{\beta} (\overline{S}^{\dagger}
h)_{\alpha} \Bigr]_{V-A} \Bigl[ (\overline{\chi}^u \overline{W}
  )_{\alpha} (\overline{W}^{\dagger}\chi^u)_{\beta} \Bigr]_{V-A},
  \nonumber \\ 
O_{2C} &=& \Bigl[ (\overline{\xi}^d W)_{\alpha} (S^{\dagger}
h)_{\alpha} \Bigr]_{V-A} \Bigl[ (\overline{\chi}^u
  \overline{W})_{\beta} (\overline{W}^{\dagger}\chi^u)_{\beta}
  \Bigr]_{V-A}, \nonumber \\
O_{3C} &=&  \sum_q \Bigl[ (\overline{\xi}^q
W S^{\dagger} \overline{S})_{\beta} (\overline{S}^{\dagger}
h)_{\alpha} \Bigr]_{V-A}  \Bigl[ (\overline{\chi}^d \overline{W}
)_{\alpha} 
(\overline{W}^{\dagger}\chi^q)_{\beta} \Bigr]_{V-A}, \nonumber \\
O_{4C} &=& \sum_q \Bigl[ (\overline{\xi}^q W)_{\alpha} (S^{\dagger}
h)_{\alpha} \Bigr]_{V-A}  \Bigl[ (\overline{\chi}^d \overline{W})_{\beta}
(\overline{W}^{\dagger} \chi^q)_{\beta} \Bigr]_{V-A}.
\label{effop}
\end{eqnarray}
Here the summation over $q$ goes over to light massless quarks. Since
we specify the direction of each 
quark, we have the original operators $O_{iR}$, obtained from
Eq.~(\ref{fullop}), in which the light-to-light current forms a meson
in the $\overline{n}^{\mu}$ direction. When a light quark in the
heavy-to-light current moves in the $\overline{n}^{\mu}$ direction and
forms a meson with an antiquark in the light-to-light current, we use
the Fierz transformation first in the full theory and match the
operators. 

Note that there are no operators $O_{5C}$ and $O_{6C}$ in SCET, which
are obtained by Fierzing $O_{5R}$ and $O_{6R}$. Neglecting color
structure and the Wilson lines, $O_{5C}$ and $O_{6C}$ have the form $-2
\overline{\xi} (1+\gamma_5) h \cdot \overline{\chi} (1-\gamma_5) \chi$,
which is identically zero at leading order in SCET because 
\begin{equation}
\overline{\chi} (1+\gamma_5)\chi = \overline{\chi} \frac{\FMslash{n}
  \FMslash{\overline{n}}}{4} (1+\gamma_5)\chi \nonumber \\
=\overline{\chi} (1+\gamma_5) \frac{\FMslash{n}
  \FMslash{\overline{n}}}{4} \chi =0,
\end{equation}
since $\FMslash{\overline{n}} \chi =0$. In the literature \cite{bbns},
the effects of the operators $O_{5C}$ and $O_{6C}$ are sometimes known
as chirally-enhanced contributions. Even though the effect is formally
suppressed, the numerical values may not be negligible. When the
equation of motion for the currents is applied, there is an
enhancement factor $m_M^2/m_q m_b$, 
where $m_M$ is a meson mass and $m_q$ is the current quark
mass. But in SCET there is simply no such operator as $O_{5C}$ and
$O_{6C}$ at leading order. This contribution is formally
suppressed in powers of $\Lambda$ in SCET, but the coefficient of
these operators can be large. In phenomenological applications, we
have to know how to treat the chirally-enhanced contributions, but
we will not consider them here.

\section{Matching and the Wilson coefficients in SCET\label{sec4}}
The Wilson coefficients of the four-quark operators in SCET can be
determined by matching the full theory onto SCET. We require
that the matrix elements of an operator in the full theory be equal to
the matrix elements of the corresponding operator in SCET. 
\begin{figure}[b]
\begin{center}
\epsfig{file=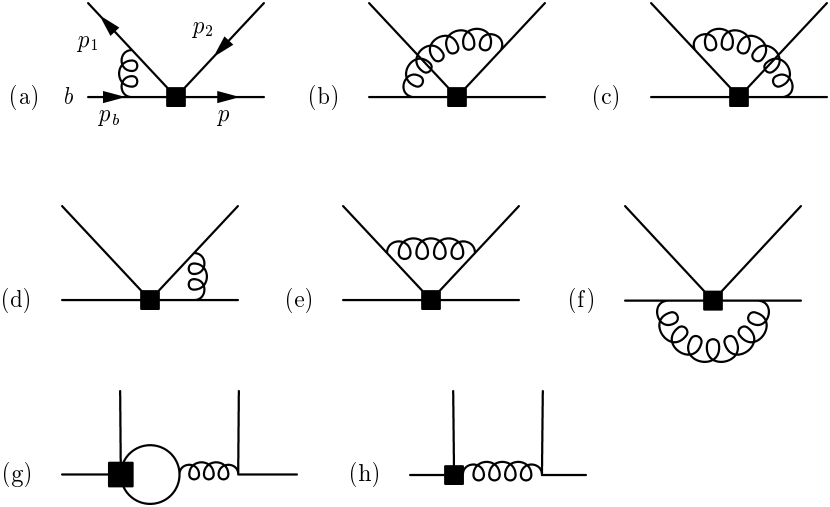, width=11.0cm}
\end{center}
\caption{Radiative corrections at one loop in the full theory. The
momenta $p_1$, $p_2$, $p$ are outgoing with $p_b = p
  +p_1+p_2$. Infrared divergences exist in diagrams (a)--(f). Diagrams
  (g) and (h) are infrared finite. In (h), the square is 
  the chromomagnetic operator $O_8$.} 
\label{fig2}
\end{figure}
We use the $\overline{\mathrm{MS}}$ scheme with naive dimensional
regularization scheme and anticommuting $\gamma_5$. All the external
particles are on the mass shell. 

The radiative corrections for the
four-quark operators in the full theory are shown in Fig.~\ref{fig2}. 
As a specific example, the amplitudes for Fig.~\ref{fig2} (a) to (d)
for the operator $O_1 = 
(\overline{d} u)_{V-A} (\overline{u} b)_{V-A}$ are given as
\begin{eqnarray}
iM_a^{(1)} &=& \frac{\alpha_s}{4\pi} (T_a)_{jk} (T_a)_{il}
(\overline{d}_j u_k)_{V-A} (\overline{u}_i b_l)_{V-A} \nonumber \\
&\times& \Bigl[\frac{1}{\epsilon_{\mathrm{UV}}} -\frac{1}{\epsilon^2}
  +\frac{2}{\epsilon} \bigl( \ln \frac{x_1 m_b}{\mu}-1\Bigr)
 -4 +2\ln \frac{m_b}{\mu} 
-2\ln^2 \frac{x_1 m_b}{\mu} \nonumber \\
&+&\frac{2-3x_1}{1-x_1}\ln x_1 - 2\mathrm{Li}_2 (1-x_1)
-\frac{\pi^2}{12}\Bigr], \nonumber \\
iM_b^{(1)} &=&  \frac{\alpha_s}{4\pi} (T_a)_{jk} (T_a)_{il}
(\overline{d}_j u_k)_{V-A} (\overline{u}_i b_l)_{V-A} \nonumber \\
&\times& \Bigl[ -\frac{4}{\epsilon_{\mathrm{UV}}}
  +\frac{1}{\epsilon^2} -\frac{2}{\epsilon} \Bigl( \ln \frac{x_2
    m_b}{\mu} -1 \Bigr) -5 + 4\ln \frac{m_b}{\mu} +2\ln^2 \frac{x_2
    m_b}{\mu} \nonumber \\
&-&\frac{2}{1-x_2} \ln x_2 +2\mathrm{Li}_2 (1-x_2)+\frac{\pi^2}{12}
  \Bigr], \nonumber \\
iM_c^{(1)} &=&  \frac{\alpha_s}{4\pi} (T_a)_{jk} (T_a)_{il}
(\overline{d}_j u_k)_{V-A} (\overline{u}_i b_l)_{V-A} \nonumber \\
&\times& \Bigl[  -\frac{4}{\epsilon_{\mathrm{UV}}}
  +\frac{2}{\epsilon^2} +\frac{2}{\epsilon} \Bigl(2-\ln
  \frac{-xx_1 m_b^2}{\mu^2}  \Bigr) -1 +\ln^2 \Bigl(
  \frac{-xx_1 m_b^2}{\mu^2} \Bigr) -\frac{\pi^2}{6} \Bigr], \nonumber
\\
iM_d^{(1)} &=&  \frac{\alpha_s}{4\pi} (T_a)_{jk} (T_a)_{il}
(\overline{d}_j u_k)_{V-A} (\overline{u}_i b_l)_{V-A} \nonumber \\
&\times& \Bigl[ \frac{1}{\epsilon_{\mathrm{UV}}} -\frac{2}{\epsilon^2}
  -\frac{2}{\epsilon} \Bigl(2 -\ln \frac{-xx_2 m_b^2}{\mu^2}  \Bigr)
  \nonumber \\
&-&8 +3 \ln \frac{-xx_2m_b^2}{\mu^2}  -\ln^2 \frac{-xx_2
   m_b^2}{\mu^2} +\frac{\pi^2}{6} \Bigr], 
\label{mat1}
 \end{eqnarray}
where $\mathrm{Li}_2 (x)$ is the dilogarithmic function. And $x$,
$x_{1,2}$ are the energy fractions given by $x=\overline{n} \cdot
p/m_b$, $x_i=n\cdot p_i/m_b$ for $i=1,2$. The prescription for $\ln
(-x_i)$ is given by $\ln (-x_i-i\epsilon) = \ln x_i -i\pi$. 
If we add all these ``nonfactorizable'' contributions, the infrared
divergence cancels and the only infrared divergence comes from the
vertex corrections of the currents [Fig.~\ref{fig2} (e) and
  (f)]. Since the vertex correction of the light-to-light current is
the same both in the full theory and in SCET, it cancels in matching.
And Fig.~\ref{fig2} (f) is given by
\begin{eqnarray}
iM_f^{(1)} &=&  \frac{\alpha_s C_F}{4\pi} (\overline{d} u)_{V-A}
(\overline{u} b)_{V-A}  \nonumber \\
&\times& \Bigl[ \frac{1}{\epsilon_{\mathrm{UV}}} -\frac{1}{\epsilon^2}
  +\frac{2}{\epsilon} \Bigl( \ln \frac{xm_b}{\mu} -1 \Bigr) -3  +2
  \ln \frac{m_b}{\mu}-2\ln^2 \frac{xm_b}{\mu} \nonumber \\
&+&\frac{2-x}{1-x} \ln x  - 2\mathrm{Li}_2 (1-x)-\frac{\pi^2}{12}
  \Bigr]. 
\end{eqnarray}

For $O_5 = (\overline{d} b)_{V-A} \sum_q (\overline{q}
q)_{V+A}$, the corresponding amplitudes are given as
\begin{eqnarray}
iM_a^{(5)} &=& \frac{\alpha_s}{4\pi} (T_a)_{jk} (T_a)_{il}
(\overline{d}_i b_l )_{V-A} (\overline{q}_j q_k)_{V+A} \nonumber \\
&\times& \Bigl[\frac{4}{\epsilon_{\mathrm{UV}}} -\frac{1}{\epsilon^2}
    +\frac{2}{\epsilon} \Bigl( \ln \frac{x_1 m_b}{\mu} -1 \Bigr) +2-4\ln
    \frac{m_b}{\mu} -2\ln^2 \frac{x_1 m_b}{\mu} \nonumber \\
&&+ \frac{2}{1-x_1} \ln x_1 -2\mathrm{Li}_2 (1-x_1) -\frac{\pi^2}{12}
    \Bigr], \nonumber \\
iM_b^{(5)} &=& \frac{\alpha_s}{4\pi} (T_a)_{jk} (T_a)_{il}
(\overline{d}_i b_l )_{V-A} (\overline{q}_j q_k)_{V+A} \nonumber \\
&\times& \Bigl[-\frac{1}{\epsilon_{\mathrm{UV}}} +\frac{1}{\epsilon^2}
  -\frac{2}{\epsilon} \Bigl( \ln \frac{x_2 m_b}{\mu} -1 \Bigr) +1-2\ln
  \frac{m_b}{\mu} +2\ln^2 \frac{x_2 m_b}{\mu} \nonumber \\
&&-\frac{2-3x_2}{1-x_2} \ln x_2 +2\mathrm{Li}_2 (1-x_2)
  +\frac{\pi^2}{12} \Bigr], \nonumber \\
iM_c^{(5)} &=& \frac{\alpha_s}{4\pi} (T_a)_{jk} (T_a)_{il}
(\overline{d}_i b_l )_{V-A} (\overline{q}_j q_k)_{V+A} \nonumber \\
&\times& \Bigl[-\frac{1}{\epsilon_{\mathrm{UV}}} +\frac{2}{\epsilon^2}
  +\frac{2}{\epsilon} \Bigl( 2-\ln \frac{-xx_1 m_b^2}{\mu^2} \Bigr) +5
  -3 \ln  \frac{-xx_1 m_b^2}{\mu^2} \nonumber \\
&&+ \ln^2  \frac{-xx_1 m_b^2}{\mu^2} -\frac{\pi^2}{6} \Bigr],
\nonumber \\
iM_d^{(5)} &=& \frac{\alpha_s}{4\pi} (T_a)_{jk} (T_a)_{il}
(\overline{d}_i b_l )_{V-A} (\overline{q}_j q_k)_{V+A} \nonumber \\
&\times& \Bigl[\frac{4}{\epsilon_{\mathrm{UV}}} -\frac{2}{\epsilon^2}
  -\frac{2}{\epsilon} \Bigl( 2-\ln \frac{-xx_2 m_b^2}{\mu^2} \Bigr) -2
  -\ln^2  \frac{-xx_2 m_b^2}{\mu^2} +\frac{\pi^2}{6}\Bigr], 
\label{mat5}
\end{eqnarray}
and $iM_f^{(5)}$ is the same as $iM_f^{(1)}$ except the Dirac
structure. The infrared divergence of the nonfactorizabie
contributions in Eq.~(\ref{mat5}) cancels, and the infrared
divergence from the vertex corrections is cancelled in
matching. 

Using the color identity
\begin{equation}
(T_a)_{jk} (T_a)_{il} =\frac{1}{2} \delta_{jl} \delta_{ik}
  -\frac{1}{2N} \delta _{jk} \delta_{il},
\end{equation}
the operators in Eq.~(\ref{mat1}) becomes $O_1/2 -O_2/(2N)$, and 
the operators in Eq.~(\ref{mat5}) becomes $O_5/2 -O_6/(2N)$. In the
radiative corrections of $O_2$, the operator becomes $C_F O_2$ due to
the color factors in Fig.~\ref{fig2}(a) and (d), while it becomes
$O_1/2-O_2/(2N)$ from Fig.~\ref{fig2}(b) and (c). Therefore the
radiative corrections 
for $O_2$, or for nonsinglet operators in general, have infrared
divergences in all the diagrams.

\begin{figure}[b]
\begin{center}
\epsfig{file=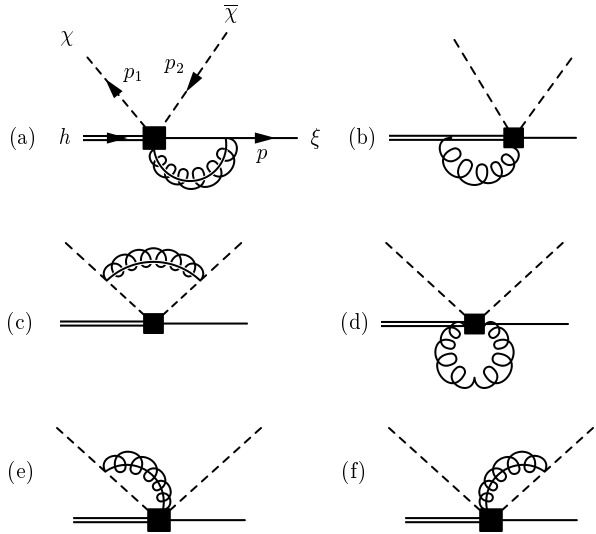, height=7.0cm}
\end{center}
\caption{Radiative corrections at one loop in SCET. Curly lines
  with a line represent collinear gluons, and curly lines represent
  soft gluons. }
\label{fig3}
\end{figure}

The corresponding radiative corrections for the four-quark operators
in SCET are shown in Fig.~\ref{fig3}. For singlet operators,
the radiative corrections exist only in the heavy-to-light current
sector [Fig.~\ref{fig3} (a), (b)], in which the infrared divergence
appearing in the calculation is exactly the same as the infrared
divergence in the full theory.  The diagram (c) is included,
but since the result is the same in the full theory, it is cancelled
in matching. 
For nonsinglet operators, we need all the diagrams in Fig.~\ref{fig3}
with additional collinear gluon interactions in the light-to-light
current sector [Fig.~\ref{fig3} (e) and (f)] and soft gluon
interactions [Fig.~\ref{fig3} (b) and (d)]. When we calculate these
diagrams, the infrared divergence is exactly the same as that
in the full theory. Therefore we can safely match the full
theory onto SCET since the infrared divergence cancels, and the
Wilson coefficients can be calculated.

In matching the theory at one loop, we calculate perturbative matrix
elements in the full and effective theories. All the long-distance
physics is reproduced in the effective theory, and the difference
between the two calculations determines the short-distance Wilson
coefficients. We treat both ultraviolet and infrared divergences using
dimensional regularization, with the final collinear quark on the mass
shell. In this case all the Feynman
diagrams in SCET are proportional to $1/\epsilon_{\mathrm{UV}}
-1/\epsilon_{\mathrm{IR}}=0$. The 
ultraviolet divergences are cancelled by the counterterms in SCET, and
all the infrared divergences cancel in the difference between the full
theory and SCET. Therefore the Wilson coefficients of
various operators in SCET can be obtained by calculating radiative
corrections in the full theory. In the full theory we also have to
consider the Feynman diagrams with fermion loops and the effect of the
chromomagnetic operator. These contributions are included in
Fig.~\ref{fig2} (g) and (h).

The Wilson coefficients are, in general, functions of the operators
$\overline{n}\cdot \mathcal{P}/m_b$, $n\cdot \mathcal{Q}/m_b$ and
$n\cdot \mathcal{Q}^{\dagger}/m_b$. The matrix elements of the
four-quark operators can be written in terms of a convolution as
\begin{eqnarray}
M&=&\int_0^1 dx dx_1 dx_2 C(x,x_1,x_2) \langle \overline{\xi} W \delta
  \Bigl(x-\frac{\overline{n} \cdot \mathcal{P}^{\dagger}}{m_b} \Bigr)
  \Gamma_1 S^{\dagger} h   \nonumber \\
&&\times \overline{\chi} \overline{W} \delta \Bigl( x_1 -\frac{n\cdot
  \mathcal{Q}^{\dagger}}{m_b} \Bigr) \Gamma_2 \delta \Bigl( x_2
  +\frac{n\cdot \mathcal{Q}}{m_b} \Bigr) \overline{W}^{\dagger} \chi
  \rangle   
\label{mael}
\end{eqnarray}
Due to the momentum conservation $x$, $x_1$ and $x_2$ satisfiy $x+x_1
+x_2 =2$. For nonleptonic decays into two light mesons at leading
order in SCET, we can set $x=1$, $x_1 =u$, and $x_2 = \overline{u}
\equiv 1-x_1$, and the matrix element can be written as
\begin{equation}
M\rightarrow \int d\eta \ C(\eta) \langle \overline{\xi} W\Gamma_1
  S^{\dagger} 
  h \cdot \overline{\chi}\overline{W} \delta (\eta -
  \mathcal{Q}_+ ) \Gamma_2 \overline{W}^{\dagger}
  \chi\rangle.
\label{fqop}
\end{equation}
where $u=x_1=1-x_2=\eta/(4E) +1/2$ with $\mathcal{Q}_+ = n\cdot
\mathcal{Q}^{\dagger} + n\cdot \mathcal{Q}$. However, we list all the
Wilson coefficients for general $x$, $x_1$ and $x_2$, which will be
useful for other decay modes, or nonleptonic $B$ decays at subleading
order in SCET.

By adding the wave function renormalization of the heavy quark, the
Wilson coefficients at next-to-leading order are given as 
\begin{eqnarray}
C_{1R}  &=& \Bigl[1-\frac{A_1}{2N} + C_F A_2 \Bigr] C_1
+\frac{A_3}{2} C_2 , \  C_{2R} = \frac{A_1}{2} C_1 +\Bigl[
  1-\frac{A_3}{2N} +C_F A_4 \Bigr] C_2, \nonumber \\
C_{3R}^p &=& \Bigl[1-\frac{A_1}{2N}  + C_F A_2 \Bigr] C_3 +\frac{A_3}{2}
C_4, \ C_{4R}^p = \frac{A_1}{2} C_3 +\Bigl[ 1-\frac{A_3}{2N} +C_F A_4
  \Bigr] C_4, \nonumber \\
C_{5R}^p &=& \Bigl[ 1-\frac{A_5}{2N} +C_F A_2 \Bigr] C_5 +\frac{A_6}{2}
C_6, \ C_{6R}^p = \frac{A_5}{2} C_5 +\Bigl[1-\frac{A_6}{2N} +C_F A_7
  \Bigr] C_6, \nonumber \\
C_{1C}&=& \Bigl[ 1-\frac{A_3}{2N} +C_F A_4 \Bigr] C_1 + \frac{A_1}{2}
C_2, \ C_{2C} = \frac{A_3}{2} C_1 + \Bigl[1-\frac{A_1}{2N} + C_F A_2
  \Bigr] C_2\nonumber \\ 
C_{3C}^p&=&\Bigl[ 1-\frac{A_3}{2N} +C_F A_4 \Bigr] C_3 +  \frac{A_1}{2}
C_4-\frac{1}{2N} C_l^p, \nonumber \\
C_{4C}^p &=&  \frac{A_3}{2} C_3 +   \Bigl[1-\frac{A_1}{2N}  + C_F A_2
  \Bigr] C_4 +\frac{1}{2}  C_l^p,
\label{wilco}
\end{eqnarray}
where $C_i$'s are the Wilson coefficients from the full theory. And
the coefficients $A_i$ to order $\alpha_s$, evaluated at $\mu=m_b$ are
given as  
\begin{eqnarray}
A_1&=& \frac{\alpha_s}{4\pi} \Bigl[ -18 + 3\ln (-x)  +\ln
  \frac{x^2}{x_1 x_2} \ln \frac{x_1}{x_2} 
+\frac{2-3x_1}{1-x_1} \ln x_1 
+\frac{1-3x_2}{1-x_2} \ln x_2 \nonumber \\
&&-2 \mathrm{Li}_2 (1-x_1 ) +2 \mathrm{Li}_2 (1-x_2)  \Bigr]
  \nonumber \\ 
A_2&=& \frac{\alpha_s}{4\pi} \Bigl[-5
  -2 \ln^2 x +\frac{2-x}{1-x} \ln x -2 \mathrm{Li}_2
  (1-x) -\frac{\pi^2}{12} \Bigr]    \nonumber \\
A_3 &=& \frac{\alpha_s}{4\pi} \Bigl[ -9  
 +\frac{2-x}{1-x} \ln x  -2 \mathrm{Li}_2 ( 1-x)+2\ln^2 x_2 
 +2\ln^2 (-x_1) -\ln^2 \frac{-x_1}{x}\nonumber \\
&& -\frac{2}{1-x_2}
  \ln x_2 +2 \mathrm{Li}_2(1-x_2) -\frac{\pi^2}{6}\Bigr],  \nonumber \\
A_4 &=& \frac{\alpha_s}{4\pi} \Bigl[-14 
  -2\ln^2 x_1 -\ln^2 (-xx_2)   +3\ln (-xx_2) 
+\frac{2-3x_1}{1-x_1} \ln x_1 \nonumber \\
&&-2\mathrm{Li}_2 (1-x_1)
  +\frac{\pi^2}{12} \Bigr],  \nonumber \\
A_5  &=&  \frac{\alpha_s}{4\pi} \Bigl[6 -3 \ln
 (- x)  +\ln \frac{x^2}{x_1 x_2} \ln \frac{x_1}{x_2}
  -\frac{1-3x_1}{1-x_1} \ln x_1 
-\frac{2-3x_2}{1-x_2} \ln x_2   \nonumber \\
&&-2 \mathrm{Li}_2 (1-x_1) +2
  \mathrm{Li}_2  ( 1-x_2) \Bigr],
  \nonumber \\
A_6 &=& \frac{\alpha_s}{4\pi} \Bigl[3 +2\ln^2 x_2 +2\ln^2 (-x_1)
  -\ln^2   \frac{-x_1}{x} 
-\frac{1-2x}{1-x} \ln x -3\ln (-x_1)\nonumber \\
&&  -\frac{2-3x_2}{1-x_2} \ln x_2 
-2 \mathrm{Li}_2 (1-x) +2 \mathrm{Li}_2 ( 1-x_2)
  -\frac{\pi^2}{6} \Bigr], \nonumber \\
A_7 &=&\frac{\alpha_s}{4\pi} \Bigl[ -2  -2\ln^2 x_1 -\ln^2 (-xx_2)
+\frac{2}{1-x_1} \ln x_1 \nonumber \\
&&-2\mathrm{Li}_2 (1-x_1) +\frac{\pi^2}{12} \Bigr]. 
\end{eqnarray}
And the contribution $C_l^p$ from fermion loops and the
chromomagnetic operator [Fig.~\ref{fig2} (g) and (h)] at $\mu=m_b$ are
given by 
\begin{eqnarray}
C_l^p&=& \frac{\alpha_s}{4\pi} \Bigl[ C_1 \Bigl( \frac{2}{3}
 -G(s_p) \Bigr) +C_3 \Bigl(
  \frac{4}{3} -G(0) -G(1) \Bigr)
  \nonumber \\
&&+C_4 \Bigl( \frac{2n_f}{3} 
  -3G(0) -G(s_c) -G(1) \Bigr) \nonumber \\
&&+C_6 \Bigl( -3G(0) -G(s_c) -G(1)
  \Bigr) -(C_5+C_8)\frac{2}{1-x_1} \Bigr], 
\label{clp}
\end{eqnarray}
where $s_p=m_p^2/m_b^2$ ($s_c = m_c^2/m_b^2$), and $G(s)$ is given by
\begin{equation}
G(s) =-4\int_0^1 dz \, z(1-z) \Bigl(s-z(1-z) (1-x_1) -i\epsilon \Bigr).
\end{equation}
We can add $C_l^p/(2N)$ in $C_3^R$ and $C_5^R$,
and $C_l^p/2$ in $C_4^R$ and $C_6^R$ with $x_1 \rightarrow x$ in
Eq.~(\ref{clp}). But in physical processes in which the 
final-state consists of color singlet mesons, the contribution of
$C_l^p$ cancels, hence omitted in Eq.~(\ref{wilco}).  

\section{Nonfactorizable spectator contributions\label{sec5}}
In nonleptonic $B$ decays, we also have to consider the spectator
quark which can interact with
the $b$ quark or other quarks forming light mesons. In this section,
we concentrate on the spectator quark interacting with the quark and
antiquark pair ($\chi$ fields) produced by the weak current, which we
call nonfactorizable spectator contributions. The spectator quark can
interact with the heavy-to-light current, which will be treated in the
next section. Here we apply the
two-step matching explicitly. The time-ordered products are
constructed in $\mathrm{SCET}_{\mathrm{I}}$, and they are evolved down
to $\mathrm{SCET}_{\mathrm{II}}$. 
  
\begin{figure}[t]
\begin{center}
\epsfig{file=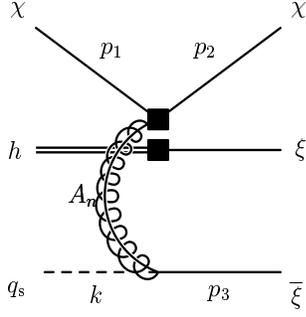, width=4.0cm}
\end{center}
\caption{A Feynman diagram for nonfactorizable spectator
contributions from the subleading operator in the light-to-light
current. The soft momentum $k$ is incoming, and $p_i$ ($i=1,2,3,4$)
are outgoing.} 
\label{fig4}
\end{figure}

Nonfactorizable spectator contributions arise from the interactions of
the gluons $A_n^{\mu}$ in the light-to-light
current with a spectator quark, which becomes a collinear quark as a
result. However the operators $O_S$ and $O_N$ at leading order
in SCET do not involve the interaction of $A_n^{\mu}$. Therefore we
take into account subleading operators which involve 
$A_n^{\mu}$ in the light-to-light current. And the Lagrangian
describing the interaction of collinear and usoft quarks begins with
$\mathcal{O} (\lambda)$ compared to the leading collinear
Lagrangian. But the propagator of the exchanged gluon is of order
$\lambda^{-2}$. Therefore the nonfactorizable spectator
contribution is of the same order as the leading
contributions from the four-quark operators. In order to evaluate
decay amplitudes at leading order in SCET, we need to include the
nonfactorizable spectator contributions, as shown in
Fig.~\ref{fig4}. Here we consider a collinear gluon $A_n^{\mu}$ from the
light-to-light current interacting with a spectator quark to produce a
collinear quark $\xi$.

The Lagrangian for the  collinear and usoft quarks at order $\lambda$
is given by \cite{pirjol,beneke1}
\begin{equation}
\mathcal{L}_{\xi q}^{(1)} = \overline{q}_{us} \Bigl[ W^{\dagger}
i\FMslash{D}_n^{\perp}W \Bigr] W^{\dagger} \xi + \mathrm{h.c.}.
\end{equation}
In $\mathrm{SCET}_{\mathrm{I}}$, there are two independent subleading
operators suppressed by $\lambda$,  in which collinear gluons
$A_n^{\mu}$ come from the current $\overline{\chi} \Gamma_2
\chi$. They are given by  
\begin{eqnarray}
O^{(1a)}_i&=&  \Bigl( (\overline{\xi}W)_{\beta} \Gamma_{1i} h_{\alpha}
 \Bigr) \Biggl\{ \Bigl(\overline{\chi} \overline{W} \frac{1}{n\cdot
 \mathcal{Q}^{\dagger}} \frac{\FMslash{n}}{2} \Bigl[ W^{\dagger}
  i \overleftarrow{\FMSlash{D}}_{n\perp} W\Bigr]\Bigl)_{\alpha}
 \Gamma_{2i} (\overline{W}^{\dagger} \chi)_{\beta}  \Biggr\} \nonumber
 \\ 
O^{(1b)}_i &=& \Bigl( (\overline{\xi}W)_{\beta} \Gamma_{1i} h_{\alpha}
 \Bigr) \Biggl\{ (\overline{\chi}
\overline{W})_{\alpha} \Gamma_{2i} \Bigl( \Bigl[ W^{\dagger}
  i\overrightarrow{\FMSlash{D}}_{n\perp} W\Bigr] \frac{\FMslash{n}}{2}
 \frac{1}{n\cdot \mathcal{Q}} \overline{W}^{\dagger} \chi
 \Bigr)_{\beta}  \Biggr\},
\label{o1ab}
\end{eqnarray}
where the index $i$ runs over all the possible forms of the operators
as shown in Eq.~(\ref{effop}). The Wilson coefficients of these
operators are 1 at lowest order in $\alpha_s$. The form of the
operators $O^{(1a,1b)}_i$ can be obtained in a straightforward manner,
but the ordering of the Wilson lines is nontrivial. It will be
explained using the auxiliary field method in Appendix B. 

In $\mathrm{SCET}_{\mathrm{I}}$, the nonfactorizable spectator
contribution is given by the matrix elements of the time-ordered
product
\begin{equation}
T^{(1)}_i = \int d^4 x T \Bigl[ O^{(1a)}_i (0) + O^{(1b)}_i (0),
 i \mathcal{L}_{\xi q}^{(1)}   (x)   \Bigr].
\end{equation}
In order to go down to $\mathrm{SCET}_{\mathrm{II}}$, we decouple the
collinear-usoft interaction using the field redefinitions \cite{bauer6} 
\begin{eqnarray}
\xi^{(0)} &=& Y^{\dagger} \xi, \ \ A_n^{(0)} = Y^{\dagger}A_n Y,
\ \ 
Y(x) = \mathrm{P} \exp \Bigl( ig \int_{-\infty}^x ds n\cdot
  A_{us} (ns) \Bigr), \nonumber \\
\chi^{(0)} &=& \overline{Y}^{\dagger} \chi, \ \ A_{\overline{n}}^{(0)}
= \overline{Y}^{\dagger}A_{\overline{n}} \overline{Y},
\ \ \overline{Y} (x) = \mathrm{P} \exp \Bigl( ig \int_{-\infty}^x ds
\overline{n} \cdot A_{us} (\overline{n}s) \Bigr).
\label{redef}
\end{eqnarray}
The collinear effective Lagrangian can be written in terms of
$\xi^{(0)}$, $\chi^{(0)}$ and $A_n^{(0)\mu}$,
$A_{\overline{n}}^{(0)\mu}$, that is, the collinear-usoft
interactions are factorized with the field redefinitions given in
Eq.~(\ref{redef}). 

Matching at $\mu_0 \sim \sqrt{m_b \Lambda}$, the small expansion
parameter changes from $\lambda\sim \sqrt{\Lambda/m_b}$ to
$\lambda^{\prime} \sim \Lambda/m_b$ with a definite power counting
procedure for any operators. Recall that we rename the usoft field
of order $p\sim m_b \lambda^2 \sim m_b
\lambda^{\prime}$ as the soft field in
$\mathrm{SCET}_{\mathrm{II}}$. In that sense 
the usoft fields become soft and the Wilson line $Y$ becomes the
Wilson line with soft gluons in $\mathrm{SCET}_{\mathrm{II}}$ but
``soft'' means the momentum of order $m_b \lambda^{\prime}$. Without
introducing addtional Wilson lines, we denote this Wilson line as $S$
and in the matching we replace $Y\rightarrow S$ without any
ambiguity. And the operators are
matched onto the operators in $\mathrm{SCET}_{\mathrm{II}}$. 

The nonfactorizable spectator contribution comes from the matrix
elements of six-quark operators. In calculating the matrix elements of
$T^{(1)}_i$, we first factorize the usoft-collinear interactions using
Eq.~(\ref{redef}) with $Y\rightarrow S$, $\overline{Y} \rightarrow
\overline{S}$, and project out color indices in such a way that the
quark bilinears ($\overline{\xi}W$, $W^{\dagger} \xi$) and 
($\overline{\chi}\overline{W}$,  $\overline{W}^{\dagger} \chi$)
forming mesons are color singlets. Since the final expression
involving the Wilson lines are nontrivial, we show explicitly how the
color projection is performed for $O^{(1a)}_i$. At order $\alpha_s$,
we will extract a collinear gluon from the factor ($W^{\dagger}
i\FMSlash{D}_{n\perp} W$) in $\mathcal{L}_{\xi
  q}$ and $O^{(1a)}_i$ to contract them. Neglecting the Dirac
structure and keeping color-dependent parts only, the time-ordered
product of $O^{(1a)}_i$ with $\mathcal{L}_{\xi q}$ has the form
\begin{eqnarray}
&&\Bigl[ ( \overline{\xi} WS^{\dagger} )_{\beta} \cdot
  h_{\alpha} \Bigr] \Bigl[ \Bigl(\overline{\chi} \overline{W}
  \overline{S}^{\dagger} S T_a S^{\dagger} \Bigr)_{\alpha} \cdot
  ( \overline{S} \overline{W}^{\dagger} \chi )_{\beta}
  \Bigr] \Bigl[ ( \overline{q}_{s} S)_{\gamma}
  (T_a)_{\gamma \delta} ( W^{\dagger}\xi )_{\delta} \Bigr]
  \nonumber \\
&=& \Bigl[( \overline{\xi} W)_{\mu} (S^{\dagger})_{\mu
  \beta} \cdot h_{\alpha} \Bigr] \Bigl[ ( \overline{\chi}
  \overline{W} )_{\nu} \Bigl( \overline{S}^{\dagger} ST_a
  S^{\dagger} \Bigr)_{\nu\alpha} \cdot (\overline{S})_{\beta \rho}
  (\overline{W}^{\dagger} \chi )_{\rho} \Bigr] \nonumber \\
&&\times \Bigl[ ( \overline{q}_{s} S)_{\gamma}
  (T_a)_{\gamma \delta} ( W^{\dagger}\xi )_{\delta} \Bigr]
  \nonumber \\
&=& \Bigl[( \overline{\xi} W)_{\mu} \cdot  h_{\alpha} \Bigr]
  \Bigl[ ( \overline{\chi} \overline{W} )_{\nu} \cdot
  (\overline{W}^{\dagger} \chi )_{\rho} \Bigr] \Bigl[ (
  \overline{q}_{s} S)_{\gamma} \cdot ( W^{\dagger}\xi
  )_{\delta} \Bigr] \nonumber \\
&&\times \Bigl( \overline{S}^{\dagger} ST_a
  S^{\dagger} \Bigr)_{\nu\alpha}(S^{\dagger}\overline{S})_{\mu \rho}
  (T_a)_{\gamma \delta} \nonumber \\
&\longrightarrow& \Bigl[ ( \overline{\xi} W )_{\beta} \cdot
  h_{\alpha} \Bigr] \Bigl[ (\overline{\chi}\overline{W} )_{\nu}
  \cdot (\overline{W}^{\dagger} \chi)_{\nu} \Bigr] \Bigl[
  (\overline{q}_{s} S)_{\gamma} \cdot (W^{\dagger} \xi)_{\beta}
  \Bigr] \nonumber \\
&&\times \frac{1}{N^2} (S^{\dagger} \overline{S})_{\mu \rho}
  (\overline{S}^{\dagger} ST_a S^{\dagger})_{\rho \alpha}
  (T_a)_{\gamma\mu} \nonumber \\
&=& \frac{C_F}{N^2} \Bigl[ ( \overline{\xi} W )_{\beta} \cdot
  (S^{\dagger}h) _{\alpha} \Bigr] \Bigl[ (\overline{\chi}\overline{W}
  )_{\nu}   \cdot (\overline{W}^{\dagger} \chi)_{\nu} \Bigr] \Bigl[
  (\overline{q}_{s} S)_{\alpha} \cdot (W^{\dagger} \xi)_{\beta}
  \Bigr],
\label{cpro}
\end{eqnarray}
where the dots denote the Dirac structure and the color projection is
performed after the arrow in Eq.~(\ref{cpro}). The ime-ordered product
with $O^{(1b)}_i$ can be projected in the same way. The time-ordered
products of the operators $O^{(1a,b)}_i$ with the 
color singlet structure vanish due to the color structure.

The matrix element of the time-ordered product $\langle
T_i^{(1)}\rangle$ at order $\alpha_s$ is given by
\begin{eqnarray}
\langle T^{(1)}_i \rangle &=& -4\pi \alpha_s \frac{C_F}{N^2} \int
d\overline{n}\cdot x \int \frac{dn\cdot k}{4\pi} e^{i n\cdot k
    \overline{n} \cdot x/2}
\nonumber \\
&\times& \Biggl\{ 
\langle \Bigl[(\overline{\xi} W)_{\beta} \Gamma_{1i} (S^{\dagger}
  h)_{\alpha} \Bigr] 
\Bigl[(\overline{\chi} \overline{W} )_{\gamma} \frac{1}{n\cdot
    \mathcal{Q}^{\dagger}} 
\frac{\FMslash{n}}{2} \gamma_{\perp}^{\mu} \Gamma_{2i}
(\overline{W}^{\dagger} \chi)_{\gamma} \Bigr] \nonumber \\
&& \times \Bigl[
  (\overline{q}_s
S)_{\alpha} (\overline{n}\cdot x) \frac{1}{n\cdot \mathcal{R}^{\dagger}}
\gamma_{\mu}^{\perp}\frac{1}{\overline{n}\cdot   \mathcal{P}}
  (W^{\dagger} \xi)_{\beta} (0) \Bigr] 
\rangle \nonumber \\
&+& \langle \Bigl[(\overline{\xi} W)_{\beta}
  \Gamma_{1i} (S^{\dagger}   h)_{\alpha} \Bigr] 
\Bigl[(\overline{\chi} \overline{W} )_{\gamma}
\Gamma_{2i}  \gamma_{\perp}^{\mu} \frac{\FMslash{n}}{2}
\frac{1}{n\cdot   \mathcal{Q}} 
(\overline{W}^{\dagger} \chi)_{\gamma} \Bigr] \nonumber \\
&&\times \Bigl[  (\overline{q}_s S)_{\alpha}(\overline{n}\cdot x)
  \frac{1}{n\cdot 
  \mathcal{R}^{\dagger}}\gamma_{\mu}^{\perp}
\frac{1}{\overline{n}\cdot 
  \mathcal{P}} (W^{\dagger} \xi)_{\beta} (0)\Bigr]
\rangle
\Biggr\}. 
\label{spec}
\end{eqnarray}

Let us consider in detail the matrix elements in Eq.~(\ref{spec}) for
different Dirac structure $\Gamma_{1i} \otimes \Gamma_{2i}$. For
simplicity we omit all the Wilson lines and the momentum operators in
the following calculation. For $\gamma_{\nu} (1-\gamma_5) \otimes
\gamma^{\nu} (1-\gamma_5)$, we can evaluate the first term in the
curly bracket in Eq.~(\ref{spec}) as 
\begin{eqnarray}
&&\langle \overline{\xi}_{\beta} \gamma_{\nu} (1-\gamma_5) h_{\alpha}
  \cdot \overline{\chi} \frac{\FMslash{n}}{2} \gamma_{\perp}^{\mu}
  \gamma^{\nu} (1-\gamma_5) \chi \cdot \overline{q}_{\alpha}
  \gamma^{\perp}_{\mu} \xi_{\beta} \rangle \nonumber \\
&&=\langle \overline{\xi}_{\beta} \gamma_{\nu}^{\perp} (1-\gamma_5)
  h_{\alpha} \cdot \overline{\chi} \frac{\FMslash{n}}{2}
  (2g_{\perp}^{\mu\nu} -\gamma_{\perp}^{\nu} \gamma_{\perp}^{\mu})
  (1-\gamma_5) \chi \cdot \overline{q}_{\alpha} \gamma^{\perp}_{\mu}
  \xi_{\beta} \rangle \nonumber \\
&&= 2\langle \overline{\xi}_{\beta} \gamma_{\perp}^{\mu} (1-\gamma_5)
  h_{\alpha} \cdot \overline{q}_{\alpha} \gamma_{\mu}^{\perp}
  \xi_{\beta} \cdot \overline{\chi} \frac{\FMslash{n}}{2} (1-\gamma_5)
  \chi \rangle \nonumber \\
&&-\langle \overline{\xi}_{\beta} \gamma_{\nu}^{\perp} (1-\gamma_5)
  h_{\alpha} \cdot \overline{\chi} \frac{\FMslash{n}}{2}
  \gamma_{\perp}^{\nu} \gamma_{\perp}^{\mu} (1-\gamma_5) \chi \cdot
  \overline{q}_{\alpha} \gamma_{\mu}^{\perp} \xi_{\beta}\rangle.
\label{matcal}
\end{eqnarray}
In the first line we can replace $\gamma_{\nu}$
by $\gamma_{\nu}^{\perp}$ and we use the Fierz
transformation to arrive at the last relation. The first term in the
last relation in Eq.~(\ref{matcal}) can be further simplified as 
\begin{eqnarray}
&&2\overline{\xi}_{\beta} \gamma_{\perp}^{\mu} (1-\gamma_5)
  h_{\alpha} \cdot \overline{q}_{\alpha} \gamma_{\mu}^{\perp}
  \xi_{\beta} \cdot \overline{\chi} \frac{\FMslash{n}}{2} (1-\gamma_5)
  \chi  \nonumber \\
&&= \frac{1}{2} \overline{\xi}_{\beta} \gamma^{\mu} (1-\gamma_5)
  h_{\alpha} \cdot \overline{q}_{\alpha} \Bigl[ \gamma_{\mu}
  (1-\gamma_5) +\gamma_{\mu} (1+\gamma_5) \Bigr] \xi_{\beta} \cdot
  \overline{\chi} \FMslash{n} (1-\gamma_5)  \chi  \nonumber \\
&&=\frac{1}{2}  \overline{\xi} \gamma^{\mu} (1-\gamma_5) \xi \cdot
  \overline{q} \gamma_{\mu} (1-\gamma_5) h  \cdot \overline{\chi}
  \FMslash{n} (1-\gamma_5)  \chi  \nonumber \\
&&= \frac{1}{4}  
  \overline{\xi} \FMslash{\overline{n}} (1-\gamma_5) \xi \cdot
  \overline{q} \FMslash{n} (1-\gamma_5) h \cdot  \overline{\chi}
  \FMslash{n} (1-\gamma_5)  \chi. 
\label{firterm}
\end{eqnarray}
The part proportional to $\gamma^{\mu}
(1+\gamma_5)$ vanishes when we apply the Fierz transformation. In the
third line we use the Fierz transformation for the product of the
first two currents. Similarly, the second term in Eq.~(\ref{matcal})
is simplified as
\begin{eqnarray}
&&-\overline{\xi}_{\beta} \gamma_{\nu} (1-\gamma_5) h_{\alpha}
  \cdot \overline{\chi} \frac{\FMslash{n}}{2} \gamma_{\perp}^{\nu}
  \gamma_{\perp}^{\mu} (1-\gamma_5) \chi \cdot \overline{q}_{\alpha}
  \gamma_{\mu}^{\perp} \xi_{\beta}  \nonumber \\
&& = -\overline{\xi}_{\beta} \gamma_{\nu} (1-\gamma_5) h_{\alpha}
  \cdot \overline{\chi} \frac{\FMslash{n}}{2} \gamma^{\nu}
 (1+\gamma_5) \gamma_{\perp}^{\mu}\chi \cdot \overline{q}_{\alpha}
  \gamma_{\mu}^{\perp} \xi_{\beta}  \nonumber \\
&&= 2 \overline{\xi}_{\beta} \gamma_{\perp}^{\mu} (1-\gamma_5)
  \chi_{\gamma} \cdot \overline{\chi}_{\gamma} \frac{\FMslash{n}}{2}
  (1-\gamma_5) h_{\alpha} \cdot \overline{q}_{\alpha}
  \gamma_{\mu}^{\perp} \xi_{\beta} \nonumber \\
&&=   \overline{\xi}_{\beta} \gamma^{\mu} (1-\gamma_5)
  \chi_{\gamma} \cdot \overline{q}_{\alpha}
  \gamma_{\mu} (1-\gamma_5) \xi_{\beta}\cdot \overline{\chi}_{\gamma}
  \frac{\FMslash{n}}{2}   (1-\gamma_5) h_{\alpha} \nonumber \\
&&=\frac{1}{2}  \overline{\xi} \FMslash{\overline{n}} (1-\gamma_5)
  \xi \cdot \overline{q}_{\alpha} \FMslash{n} (1-\gamma_5)
  \chi_{\gamma} \cdot \overline{\chi}_{\gamma} (1+\gamma_5)
  \frac{\FMslash{n}}{2} h_{\alpha} \nonumber \\
&&=- \frac{1}{4} \overline{\xi} \FMslash{\overline{n}} (1-\gamma_5)\xi
  \cdot \overline{q} \FMslash{n} (1-\gamma_5) h \cdot \overline{\chi}
  \FMslash{n} (1-\gamma_5) \chi,
\label{secterm}
\end{eqnarray}
where the  Fierz transformations are applied successively.
When we add Eqs.~(\ref{firterm}) and
(\ref{secterm}), they cancel. The second term in Eq.~(\ref{spec}) is
given by Eq.~(\ref{secterm}) with an opposite sign. As a result, 
for $\Gamma_{1i} \otimes \Gamma_{2i}= \gamma_{\nu} (1-\gamma_5)
\otimes \gamma^{\nu} (1-\gamma_5)$, only the second term in
Eq.~(\ref{spec}) contributes and the matrix element is given as
\begin{eqnarray}
&&\frac{1}{4} \langle \overline{\xi} W \frac{1}{\overline{n} \cdot
  \mathcal{P}}  W^{\dagger}\FMslash{\overline{n}}
  (1-\gamma_5)  \xi \rangle \langle \overline{q}_s S
  \frac{1}{n\cdot   \mathcal{R}^{\dagger}} 
S^{\dagger} \FMslash{n} (1-\gamma_5) h\rangle \nonumber \\
&&\times \langle \overline{\chi}
\overline{W} \frac{1}{n\cdot \mathcal{Q}} \overline{W}^{\dagger}
\FMslash{n} (1-\gamma_5) \chi\rangle.
\label{me1}
 \end{eqnarray}
Note that the matrix element of the six-quark operators in
Eq.~(\ref{me1}) is written in a factorized form. It is because a
collinear or a soft gluon in one sector cannot interact with the
particles in the other sector. That type of interaction makes the
intermediate states off the mass shell by  $\sim m_b$ or
$\sqrt{m_b \Lambda}$. These off-shell
modes are already taken care of in the Wilson
coefficients or in the jet functions. Therefore the six-quark
operators describing the spectator interactions are factorized and the
matrix elements of the six-quark operators can be computed by the
products of the matrix elements of each current.

For $\gamma_{\nu} (1-\gamma_5) \otimes
\gamma^{\nu} (1+\gamma_5)$, only the first term in Eq.~(\ref{spec})
contributes and the matrix element is given by
\begin{eqnarray}
&&\frac{1}{4} \langle \overline{\xi} W \frac{1}{\overline{n} \cdot
  \mathcal{P}}  W^{\dagger} \FMslash{\overline{n}}  (1-\gamma_5) 
 \xi \rangle \langle \overline{q}_s S
  \frac{1}{n\cdot \mathcal{R}^{\dagger}} 
S^{\dagger} \FMslash{n} (1-\gamma_5) h \rangle \nonumber \\
&&\times\langle \overline{\chi}
\overline{W} 
\frac{1}{n\cdot \mathcal{Q}^{\dagger}} \overline{W}^{\dagger} \FMslash{n}
(1+\gamma_5)\chi\rangle.
\label{me2}
 \end{eqnarray}
And for $(1-\gamma_5)\otimes (1+\gamma_5)$, the matrix element is
given as
\begin{eqnarray}
&&\frac{1}{16} \langle \overline{\xi} W \frac{1}{\overline{n} \cdot
    \mathcal{P}} W^{\dagger}\FMslash{\overline{n}} (1+\gamma_5) \xi
    \rangle \langle \overline{q}_s S 
\frac{1}{n\cdot   \mathcal{R}^{\dagger}} 
S^{\dagger} \FMslash{n} \gamma_{\perp}^{\mu} (1-\gamma_5) h \rangle
\nonumber \\ &&\times\langle \overline{\chi}
\overline{W} 
\Bigl(\frac{1}{n\cdot \mathcal{Q}^{\dagger}} -\frac{1}{n\cdot
    \mathcal{Q}}\Bigr)  \overline{W}^{\dagger}
\FMslash{n} \gamma_{\mu}^{\perp} (1+\gamma_5)\chi\rangle.
\label{me3}
 \end{eqnarray}
As can be clearly seen in the final expressions in
Eqs.~(\ref{me1})--(\ref{me3}), the gluons $A_n^{\mu}$,
$A_{\overline{n}}^{\mu}$, and $A_s^{\mu}$ can be exchanged only inside
each meson. This is true at higher orders of $\alpha_s$. Though the
Wilson coefficients can be different, the structure of the operator is
the same to all orders in $\alpha_s$. Therefore the nonfactorizable
spectator contribution is factorized to all orders in $\alpha_s$. 

Let us calculate the matrix element $T^{(1)}_i$ explicitly for
$\Gamma_{1i} \otimes \Gamma_{2i} =\gamma_{\nu} (1-\gamma_5) \otimes
\gamma^{\nu} (1-\gamma_5)$. It is given by
\begin{eqnarray}
\langle T^{(1)}_i \rangle &=& -4\pi \alpha_s \frac{C_F}{N^2} \int
d\overline{n} \cdot x \int \frac{dn\cdot k}{4\pi} \frac{e^{in\cdot k
\overline{n}\cdot x/2}}{n\cdot k n\cdot p_2\overline{n} \cdot p_3}
\nonumber \\
&\times& \frac{1}{4} \langle M_1| \overline{\xi} W
\FMslash{\overline{n}} (1-\gamma_5) W^{\dagger} \xi |0\rangle \langle
M_2| \overline{\chi} \overline{W} \FMslash{n} (1-\gamma_5)
\overline{W}^{\dagger} \chi|0\rangle \nonumber \\
&\times& \langle 0| \overline{q}_s S (\overline{n} \cdot x) \FMslash{n}
(1-\gamma_5) S^{\dagger} h (0)|B\rangle,
\end{eqnarray}
where we integrate out $n\cdot x$ and $x_{\perp}$ explicitly. The
matrix element involving the $B$ meson can be calculated as
\begin{eqnarray}
&&\langle 0| \overline{q}_s S (\overline{n} \cdot x) \FMslash{n}
(1-\gamma_5) S^{\dagger} h |B\rangle  
=\int dr_+ e^{-ir_+ \overline{n} \cdot x/2} \mathrm{Tr} \ \Bigl[
    \Psi_B (r_+) \FMslash{n} (1-\gamma_5) \Bigr] \nonumber \\
&&=-\frac{if_B m_B}{4} \int dr_+  e^{-ir_+ \overline{n} \cdot x/2}
  \mathrm{Tr}\ \Bigl[ \frac{1+\FMslash{v}}{2} \FMslash{\overline{n}}
    \gamma_5 \FMslash{n} (1-\gamma_5) \Bigr] \phi_B^+ (r_+) \nonumber
  \\
&&= -if_B m_B \int dr_+  e^{-ir_+ \overline{n} \cdot x/2}
  \phi_B^+ (r_+),
\end{eqnarray}
where the leading-twist $B$ meson light-cone wave function is defined
through the projection of the $B$ meson as \cite{grozin,feldmann}
\begin{equation}
\Psi_B (r_+) =-\frac{if_B m_B}{4} \Bigl[ \frac{1+\FMslash{v}}{2}
    \Bigl( \FMslash{\overline{n}} \phi_B^+ (r_+) + \FMslash{n}
    \phi_B^- (r_+) \Bigr) \gamma_5 \Bigr].
\end{equation}
And the light-cone wave function for the light mesons is defined as
\begin{eqnarray}
&&\langle M_2| \overline{\chi}\overline{W} \FMslash{n} \gamma_5
  \delta (\eta -   \mathcal{Q}_+ ) 
  \overline{W}^{\dagger}   \chi\rangle |0\rangle \nonumber \\
&& =-if_{M2} 2E \int_0^1
  du \delta [\eta -(4u-2)E] \phi_{M2} (u).
\end{eqnarray}

For $\gamma_{\nu} (1-\gamma_5) \otimes \gamma^{\mu} (1\mp \gamma_5)$,
the matrix element which is given by
\begin{eqnarray}
\langle T^{(1)}_i\rangle &=& \frac{iC_F \pi \alpha_s}{N^2}f_B f_{M1}
f_{M2} m_B \nonumber \\
&&\times \int dr_+ \frac{\phi_B^+ (r_+)}{r_+} \int du
\frac{\phi_{M_1}(u)}{u} \int dv \frac{\phi_{M2} (v)}{v}.
\label{specc}
\end{eqnarray}
For $(1-\gamma_5) \otimes (1+\gamma_5)$, the matrix element is zero if
we use the leading-twist $B$ meson wave function because
\begin{eqnarray}
\langle 0| \overline{q}_s \FMslash{n} \gamma_{\perp}^{\mu}
(1+\gamma_5) h| B\rangle &=& -\frac{if_B m_B}{4} \mathrm{tr} \ \Bigl[
  \FMslash{n} \gamma_{\perp}^{\mu} (1+\gamma_5)
  \frac{1+\FMslash{v}}{2} \Bigl( \FMslash{\overline{n}} \phi_B^+
  +\FMslash{n} \phi_B^- \Bigr) \gamma_5 \Bigr] \nonumber \\
&=& -\frac{if_B m_B}{8} \phi_B^+ \mathrm{tr} \ \FMslash{v}
  \FMslash{\overline{n}} \FMslash{n} \gamma_{\perp}^{\mu} =0.
\end{eqnarray}
If we use higher-twist wave function for the $B$ meson, there may be
nonzero contributions, but this is expected to be suppressed. 

\section{Spectator contribution to the form factor\label{sec6}}
In Section~\ref{sec5}, we have considered the nonfactorizable spectator
contributions arising from the subleading operators in which
we include only the subleading part from the light-to-light
current. However, there is also a subleading operator coming from the
heavy-to-light current, but this contributes to the heavy-to-light
form factor. It is considered first in Ref.~\cite{form}, and we
discuss in detail here in the context of nonleptonic $B$ decays.

The operators in $\mathrm{SCET}_{\mathrm{I}}$,
which contribute to the form factor, are given by
\begin{eqnarray}
J^{(0)}_i &=& (\overline{\xi} W \Gamma_{1i} h) (\overline{\chi}
\overline{W} \Gamma_{2i} \overline{W}^{\dagger}\chi), \nonumber \\
J^{(1a)}_i &=& \Bigl(\overline{\xi} W
(W^{\dagger} i\overleftarrow{\FMSlash{D}}_{n\perp} W)
  \frac{\Gamma_{1i}}{\overline{n}\cdot \mathcal{P}^{\dagger}}
  h\Bigr)(\overline{\chi} \overline{W} \Gamma_{2i}
  \overline{W}^{\dagger}   \chi), \nonumber \\
 J^{(1b)}_i &=&\Bigl( \overline{\xi} W
(W^{\dagger} i\overrightarrow{\FMSlash{D}}_{n\perp} W)
  \frac{\Gamma_{1i}}{m_b} h\Bigr) (\overline{\chi} \overline{W}
  \Gamma_{2i} \overline{W}^{\dagger} \chi).
\label{fsubop}
\end{eqnarray}
Here we list only singlet operators. The nonsinglet operators give the
same matrix elements as the singlet operators with the color
suppression factor  $1/N$. We will consider only singlet operators
from now on.  

\begin{figure}[t]
\begin{center}
\epsfig{file=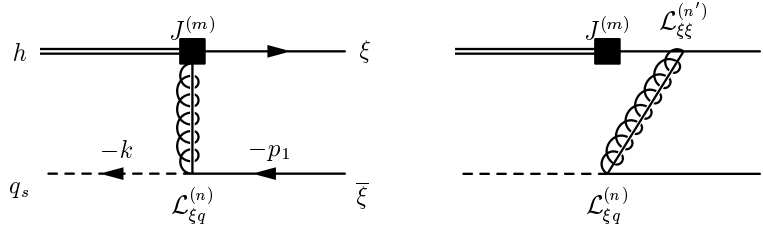, width=10.0cm}
\end{center}
\caption{Tree-level graphs in $\mathrm{SCET}_{\mathrm{I}}$ for the
  spectator contribution to the heavy-to-light form factor. The first
  diagram contributes to $T_{1,2,4}$, and the second diagram
  contributes to $T_{0,1,3,4,5,6}$.}
\label{fig5}
\end{figure}

The interaction of collinear and usoft quarks is
given by \cite{form}
\begin{eqnarray} 
\mathcal{L}_{\xi q}^{(1)} &=& ig \overline{\xi}
\frac{1}{i\overline{n}\cdot D_n} \FMSlash{B}_{\perp}^n W
  q_{us} +\mathrm{h. c.}, \ \mathcal{L}_{\xi q}^{(2a)} =ig
  \overline{\xi} \frac{1}{i\overline{n} \cdot D_n} \FMSlash{M}
  Wq_{us} +\mathrm{h.c.}, \nonumber \\
\mathcal{L}_{\xi q}^{(2b)} &=& ig \overline{\xi} \frac{\FMslash{n}}{2}
i\FMSlash{D}_{\perp}^n \frac{1}{(i\overline{n} \cdot D_n)^2}
\FMSlash{B}_{\perp}^n W q_{us} +\mathrm{h.c.},
\end{eqnarray}
where
\begin{equation}
ig \FMSlash{B}_{\perp}^n = [i\overline{n}\cdot D_n,
  i\FMSlash{D}_{\perp}^n], \ ig \FMSlash{M} =[i\overline{n} \cdot D^n,
  i\FMSlash{D}^{us} +\frac{\FMslash{\overline{n}}}{2} gn\cdot
  A_n].
\end{equation}
At leading order in SCET, the relevant time-ordered products are given
as 
\begin{eqnarray}
T_{0i}^F &=&\int d^4 x T[J^{(0)}_i (0) i\mathcal{L}_{\xi q}^{(1)} (x)
], \ T_{1i}^F = \int d^4 x T[J^{(1a)}_i (0) i\mathcal{L}_{\xi q}^{(1)}
  (x)], \nonumber \\ 
T_{2i}^F &=&
\int d^4 x T[J^{(1b)}_i (0) i\mathcal{L}_{\xi q}^{(1)} (x) ], \
T_{3i}^F = \int d^4 x T[J^{(0)}_i (0) i\mathcal{L}_{\xi q}^{(2b)} (x)
], \nonumber \\ 
T_{4i}^{NF} &=& \int d^4 x T[J^{(0)}_i (0) i\mathcal{L}_{\xi q}^{(2a)}
  (x) ], \nonumber  \\   
T_{5i}^{NF} &=& \int d^4 x d^4 y T[J^{(0)}_i (0) i\mathcal{L}_{\xi
   \xi}^{(1)} (x)  i\mathcal{L}_{\xi q}^{(1)} (y) ], \nonumber \\
T_{6i}^{NF}
&=& \int d^4 x d^4 y T[J^{(0)}_i (0) i\mathcal{L}_{cg}^{(1)} (x)
  i\mathcal{L}_{\xi q}^{(1)} (y) ], 
\label{formtime}
\end{eqnarray}
where $\mathcal{L}_{\xi \xi}^{(1)}$ is the leading collinear
Lagrangian and $\mathcal{L}_{cg}^{(1)}$ is the subleading gluon
Lagrangian, which can be found in Refs.~\cite{form,pirjol}.
The Feynman diagrams at lowest order in $\alpha_s$ for the
time-ordered products are shown in Fig.~\ref{fig5} schematically.
Compared to the case of the nonfactorizable spectator interactions,
the leading-order heavy-to-light current $J^{(0)}_i$ contains
$A_n^{\mu}$, therefore it contributes to the heavy-to-light form 
factor starting from the leading order. 

As suggested in Ref.~\cite{form}, we absorb the nonfactorizable parts
$T_{ki}^{NF}$ ($k=4,5,6$) into the definition of the soft
nonperturbative effects for the form factors at this order. Among the
factorizable contributions $T_{li}^F$, only $T_{2i}^F$ is nonzero at
order $\alpha_s$. For $\Gamma_{1i} \otimes \Gamma_{2i} = \gamma_{\nu}
(1-\gamma_5) \otimes \gamma^{\nu} (1-\gamma_5)$, the matrix element of
$T_{2i}^F$ is given by 
\begin{eqnarray}
\langle T_{2i}^F \rangle &=&\langle M_1 | i\int d^4 x T [J^{(1b)}_i
  (0),   \mathcal{L}^{(1)}_{\xi q} 
  (x)]|B \rangle \nonumber \\
&=&\frac{g^2}{4\pi} \frac{1}{m_b \overline{n} \cdot p_1} \int
  d\overline{n} \cdot x \int \frac{dn\cdot k}{n\cdot k} e^{in\cdot k
  \overline{n} \cdot x/2} \nonumber \\
 &\times& \frac{C_F}{4N} \langle \overline{\xi} W \FMslash{\overline{n}}
  (1-\gamma_5) W^{\dagger} \xi \cdot \overline{\chi} \overline{W}
  \FMslash{n}   (1-\gamma_5) \overline{W}^{\dagger} \chi 
  \cdot \overline{q}_s S (\overline{n} \cdot x) \FMslash{n}
  (1-\gamma_5) S^{\dagger} h \rangle \nonumber \\
&=& \frac{\alpha_s}{4\pi} \frac{4\pi^2 C_F}{N}  if_{M_1} f_{M2} f_B
  \frac{2E}{m_b} m_B \int du \frac{\phi_{M_1} (u)}{u} \int
  \frac{dr_+}{r_+} \phi_B^+ (r_+), 
\label{t1mat}
\end{eqnarray}
where we use Fierz transformations successively. For $\Gamma_{1i}
\otimes \Gamma_{2i} =\gamma_{\nu} (1-\gamma_5) \otimes \gamma^{\nu}
(1+\gamma_5)$, we obtain the same result as in Eq.~(\ref{t1mat}). For
$(1+\gamma_5) \otimes (1-\gamma_5)$, it vanishes.

The form factors for $B$ decays into light pseudoscalar mesons are
defined as
\begin{eqnarray}
\langle P(p)| \overline{q} \gamma^{\mu} b | \overline{B}
(p_B)\rangle &=& f_+ (q^2) \Bigl[ p_B^{\mu} + p^{\mu}
  -\frac{m_B^2-m_P^2}{q^2} q^{\mu} \Bigr] \nonumber \\
&& +f_0 (q^2) \frac{m_B^2-m_P^2}{q^2} q^{\mu},
\end{eqnarray}
where $q^{\mu}= p_B^{\mu} -p^{\mu}$. In SCET, the form factor at order
$\alpha_s$ is given by  
\begin{eqnarray}
f_+ (0) &=& \pi \alpha_s\frac{C_F}{N} \frac{f_{M1}f_B m_B}{4E^2}
\frac{2E}{m_b}  \int dx dr_+ \frac{\alpha_s (\mu)}{xr_+} \phi_{M1}
(x,\mu)   \phi_B^+ (r_+,\mu) \nonumber \\
&&+ (1+K) \zeta (\mu_0,\mu),
\label{fplus}
\end{eqnarray}
where $\zeta (\mu_0,\mu)$ is a nonperturbative function introduced in
Ref.~\cite{charles} in large-energy effective theory
\cite{grinstein}. A similar nonperturbative function is
introduced in Refs.~\cite{bauer2,chay1}, based on SCET. The procedure in
obtaining independent nonperturabtive functions in the form factor is
different in the large-energy effective theory and in SCET, but the
number of independent nonperturbative functions is the same.
And $K$ at $\mu=m_b$ is given by
\begin{equation}
K=-\frac{\alpha_s C_F}{4\pi} \Bigl(6+\frac{\pi^2}{12}\Bigr). 
\end{equation}
The expression for $f_+$ coincides with the result in Ref.~\cite{form}
with $m_B=2E$ at order $\alpha_s$, and $K$ is calculated in
Refs.~\cite{bauer2,chay1}.

\section{Application to $\overline{B} \rightarrow \pi\pi$
  decays\label{sec7}} 
As an application, we consider the decay amplitudes for $\overline{B}
\rightarrow \pi \pi$, which can be written as
\begin{equation}
\langle \pi \pi | H_{\mathrm{eff}} |\overline{B}\rangle
=\frac{G_F}{\sqrt{2}} \sum_{p=u,c} V_{pb} V_{pd}^* \langle \pi \pi |
\mathcal{A}_p |\overline{B}\rangle,
\end{equation}
where the operators $\mathcal{A}_p$ are given by
\begin{eqnarray}
\mathcal{A}_p &=& a_1^p  \Bigl[(\overline{\xi}^u h)_{V-A} 
(\overline{\chi}^d \chi^u )_{V-A} \Bigr] + a_2^p \Bigl[ (\overline{\xi}^d
h)_{V-A} (\overline{\chi}^u \chi^u )_{V-A}\Bigr] \nonumber \\
&&+a_3^p \Bigl[ (\overline{\xi}^d h)_{V-A}
(\overline{\chi}^q \chi^q )_{V-A} \Bigr] + a_4^p \Bigl[ (\overline{\xi}^q
h)_{V-A} (\overline{\chi}^d \chi^q )_{V-A} \Bigr] \nonumber \\ 
&&+a_5^p \Bigl[(\overline{\xi}^d h)_{V-A}
(\overline{\chi}^q \chi^q )_{V+A} \Bigr].
\end{eqnarray}
The notation $a_i^p [\mathcal{O}]$ means that $a_i^p$ are the sum of
the amplitudes initiated by the operator $\mathcal{O}$. They include the
contribution from the operator itself, and the spectator
contributions. That is,  $a_i^p [\mathcal{O}]$ can be written as
\begin{equation}
a_i^p [\mathcal{O}] = T_i^p + N_i^p + F_i^p,
\end{equation}
where $T_i^p$ is the contribution from the four-quark operators,
$N_i^p$ is the nonfactorizable spectator contribution, and $F_i^p$ is
the spectator contribution for the heavy-to-light form
factor with $a_1^c=a_2^c=0$. 

Before we present the decay amplitudes for $\overline{B} \rightarrow
\pi \pi$ at order $\alpha_s$ explicitly, we show the amplitudes
$a_i^p$ in general to all orders in $\alpha_s$, which are derived from
SCET.  The form of $T_i^p$ from Eq.~(\ref{fqop}) can be written as 
\begin{eqnarray}
T_i^p&=& \int d\eta \ C_{\mathrm{eff},i}^{T} (\eta, \mu_0, \mu)
 \nonumber  \\ 
&&\times \langle
  \overline{\xi} W\gamma_{\mu}  (1-\gamma_5)   S^{\dagger}   h \cdot
  \overline{\chi}\overline{W} 
  \delta (\eta -   \mathcal{Q}_+ ) \gamma^{\mu} (1\mp \gamma_5)
  \overline{W}^{\dagger}   \chi\rangle \nonumber \\
&=& \pm if_{M2} 2E \int_0^1 du C_{\mathrm{eff},i}^{T} (u,\mu_0 ,\mu)
 \phi_{M2} (u, \mu)
  \langle M_1 | 
\overline{\xi} W \frac{\FMslash{\overline{n}}}{2} (1-\gamma_5)
  S^{\dagger} h |\overline{B} \rangle \nonumber \\
&=& \pm im_B^2  f_{M2} \int_0^1 du \zeta (\mu_0, \mu)
 C_{\mathrm{eff},i}^{T} (u, \mu_0,\mu)   \phi_{M2} (u,\mu), 
\label{gen1}
\end{eqnarray}
where the upper (lower) sign corresponds to $i=1,\cdots, 4$ ($i=5$). 
The effective Wilson coefficients $C_{\mathrm{eff},i}^{T}$ are evaluated
at $\mu =m_b$ and evolved down to $\mu=\mu_0=\sqrt{m_b \Lambda}$ to
be matched onto $\mathrm{SCET}_{\mathrm{II}}$. Then they evolve down
to the scale $\mu$, where the matrix elements are evaluated.

The matrix element of the heavy-to-light current is given by
\cite{charles,chay1,bauer2}
\begin{equation}
\langle M_1 |\overline{\xi} W \FMslash{\overline{n}}
(1-\gamma_5)   S^{\dagger} h |\overline{B} \rangle = 2m_B \zeta.
\label{bpi}
\end{equation}
The nonperturbative parameter $\zeta$ is matched at $\mu=\mu_0$
and it evolves down to $\mu$, and the wave functions are evaluated at
$\mu$. Note that we do not include the radiative corrections in
$\zeta$, since they are taken into account either in the Wilson
coefficients or in the spectator contributions. 

The spectator contribution $N_i^p$ involves the time-ordered products,
which can be written as
\begin{eqnarray}
T[O^{(1a)}_i (x) + O^{(1b)}_i (x), i\mathcal{L}_{\xi q}^{(1)} (0) ]
&=& \delta \Bigl( 
\frac{n\cdot x}{2} \Bigr) \delta^2 (x_{\perp}) \int d\overline{\eta}
d\eta  \int dr_+ e^{ir_+ \overline{n} \cdot x/2} \nonumber
\\
&\times& J^N_i( \eta, \overline{\eta}, r_+)
\mathcal{O}_i (\eta,\overline{\eta},r_+),
\end{eqnarray}
where $J^N_i$ are the jet functions, which are obtained by matching
$\mathrm{SCET}_{\mathrm{I}}$ onto $\mathrm{SCET}_{\mathrm{II}}$. The
index $i$ denotes all the possible operators for the decay. And
the operator $\mathcal{O}_i$ is given by
\begin{eqnarray}
\mathcal{O}_i  (\eta,\overline{\eta},r_+)&=& \overline{\xi} W
\Gamma_{1i} \delta (\eta -\mathcal{P}_+) W^{\dagger} \xi \cdot
\overline{\chi} \overline{W} \Gamma_{2i} \delta (\overline{\eta}
-\mathcal{Q}_+) \overline{W}^{\dagger} \chi \nonumber \\
&&\times \overline{q}_s \FMslash{n} (1-\gamma_5) \delta
(\mathcal{R}^{\dagger} -r_+) S^{\dagger} h.
\end{eqnarray}
Therefore $N_i^p$ can be written as
\begin{eqnarray}
N_i^p &=& \int d^4 x C_{\mathrm{eff},i}^{N} T[O^{(1a)}_i (x) +
  O^{(1b)}_i (x), i\mathcal{L}_{\xi q}^{(1)} (0) ] \nonumber \\ 
&=&\int du dv dr_+ 
C_{\mathrm{eff},i}^{N} ( \mu_0, \mu) J^N_i (u,v,r_+, \mu_0,
\mu) N_i f_B f_{M1} f_{M_2} \nonumber \\
&&\times \phi_{M1} (u, \mu) \phi_{M2} (v,\mu) \phi_B^+ (r_+,\mu),  
\label{gen2}   
\end{eqnarray}
where $ C_{\mathrm{eff},i}^{N}$ are the effective Wilson coefficients,
and the $\mu$ dependence is shown explicitly. Here $N_i$ are the
normalization constants when we evaluate the matrix elements of
$\mathcal{O}_i$. Since $\overline{n} \cdot
p_{M1} = n\cdot p_{M2} =2E$, the variables $u$ and $v$ satisfy the
relations
\begin{equation}
u= \frac{\eta}{4E}+\frac{1}{2} , \ \ v = \frac{\overline{\eta}}{4E} 
+\frac{1}{2}.
\end{equation}

The factorizable spectator contribution to the form factor can be
written similarly as
\begin{eqnarray}
F_i^p &=& \int dudv dr_+ C_{\mathrm{eff},i}^F (\mu_0, \mu)
J_i^F (u,v, r_+,\mu_0,\mu) N_i f_B f_{M1} f_{M2} \nonumber \\
&&\times \phi_{M1} (u, \mu) \phi_{M2} (v,\mu) \phi_B^+ (r_+,\mu),     
\label{gen3}
\end{eqnarray}
where $ C_{\mathrm{eff},i}^F$ are the effective Wilson coefficients, 
and $J_i^F$ are the jet functions obtained from matching
$\mathrm{SCET}_{\mathrm{I}}$ onto $\mathrm{SCET}_{\mathrm{II}}$. The
spectator interactions $N_i^p$ in Eq.~(\ref{gen2}) and $F_i^p$ in
Eq.~(\ref{gen3}) are factorized into the short-distance and the
long-distance parts to all orders in $\alpha_s$, and the convolution
integrals are finite.

Now let us calculate the decay amplitudes for $\overline{B}
\rightarrow \pi \pi$ at order $\alpha_s$, based on the general
expressions on $T_i^p$, $N_i^p$ and  $F_i^p$. The contributions
$T_i^p$ are obtained by the convolutions of the 
following effective Wilson coefficients 
\begin{eqnarray}
C_{\mathrm{eff},1}^T &=& C_{1R} +\frac{C_{2R}}{N}, \
C_{\mathrm{eff},2}^T = C_{2C} 
+\frac{C_{1C}}{N}, \ C_{\mathrm{eff},3}^T = C_{3R} +\frac{C_{4R}}{N},
\nonumber \\ C_{\mathrm{eff},4}^{Tp} &=& C_{4C}^p
+\frac{C_{3C}^p}{N}, \ C_{\mathrm{eff},5}^{Tp}  = C_{5R}
+\frac{C_{6R}}{N},
\end{eqnarray}
with the hadronic matrix elements of the four-quark operators.
For the decay amplitudes at leading order, we can
put $x=1$ and let $u=x_1$ and $\overline{u} = x_2 = 1-u$, and the
amplitudes $T_i^p$'s, evaluated at $\mu=m_b$ are given as
\begin{eqnarray}
T_1^p &=& im_B^2 \zeta f_{\pi} \Bigl[\Bigl(C_1 +\frac{C_2}{N}\Bigr) 
  (1+K) +\frac{\alpha_s}{4\pi} \frac{C_F}{N} C_2 F \Bigr], \nonumber 
  \\ 
T_2^p &=&im_B^2 \zeta f_{\pi} \Bigl[
\Bigl(C_2 +\frac{C_1}{N}\Bigr)  (1+K) 
+\frac{\alpha_s}{4\pi} \frac{C_F}{N} C_1 F \Bigr],\nonumber \\
T_3^p &=&im_B^2 \zeta f_{\pi} \Bigl[ \Bigl(C_3 +\frac{C_4}{N}\Bigr)
  (1+K) +\frac{\alpha_s}{4\pi} \frac{C_F}{N} C_4 F \Bigr], \nonumber
  \\ 
T_4^p  &=& im_B^2 \zeta f_{\pi} \Biggl\{ \Bigl[\Bigl(C_4
  +\frac{C_3}{N}\Bigr)  (1+K)  +\frac{\alpha_s}{4\pi} \frac{C_F}{N}
  \Bigl[ C_3 F   + C_1 \Bigl( \frac{2}{3} -G(s_p) \Bigr) \nonumber \\ 
&+&C_3\Bigl( \frac{4}{3}  -G(0)  -G(1) \Bigr) 
+ C_4 \Bigl( \frac{2n_f}{3}  -3G(0)  -G(s_c) -G(1) \Bigr) \nonumber \\
&+& C_6 \Bigl(  -3G(0)
  -G(s_c) -G(1) \Bigr) +G_{\pi,8} (C_5+C_8) \Bigr] \Biggr\}, \nonumber \\
T_5^p &=&-im_B^2 \zeta f_{\pi} \Bigl[
\Bigl(C_5 +\frac{C_6}{N}\Bigr)  (1+K) 
+\frac{\alpha_s}{4\pi} \frac{C_F}{N} C_6 (-F-12) \Bigr],
\end{eqnarray}
where $C_F =(N^2-1)/(2N)$, $N=3$, $n_f=5$. And $F$ is given as
\begin{eqnarray}
F&=& -18 +f_{\pi}^{\mathrm{I}}, \ \ 
f_{\pi}^{\mathrm{I}} = \int_0^1 du g(u) \phi_{\pi} (u), \ \ G_{\pi,8}
= \int_0^1 du \frac{-2}{1-u} \phi_{\pi} (u), \nonumber \\ 
g(u) &=& 3 \frac{1-2u}{1-u} \ln u -3i\pi \nonumber \\
&&-\Bigl[ 2 \mathrm{Li}_2 (1-u)
+\frac{1-3u}{1-u} \ln u +\ln^2 u -(u\leftrightarrow \overline{u})
  \Bigr].
\end{eqnarray}

The nonfactorizable spectator contributions $N_i^p$ are given by
Eq.~(\ref{specc}) as 
\begin{equation}
N_i^p =i \frac{C_F}{N^2} \pi \alpha_s f_B f_{\pi}^2 m_B
C_{\mathrm{eff},i}^N  
\int dr_+ \frac{\phi_B^+ (r_+)}{r_+} \Bigl( \int_0^1
  \frac{\phi_{\pi} (u)}{u} \Bigr)^2,
\end{equation}
where the effective Wilson coefficients are given by
\begin{equation}
C_{\mathrm{eff},1}^N=C_2, \ C_{\mathrm{eff},2}^N = C_1, \
C_{\mathrm{eff},3}^N = C_4, \ C_{\mathrm{eff},4}^N = C_3, \
C_{\mathrm{eff},5}^N  = C_6.
\end{equation}
The spectator contributions to the heavy-to-light form factor $F_i^p$
are given as
\begin{equation}
F_i^p = i \frac{C_F}{N} \pi \alpha_s f_B f_{\pi}^2 m_B
C_{\mathrm{eff},i}^F  \int du\frac{\phi_{\pi} (u)}{u} \int dr_+
\frac{\phi_B^+ (r_+)}{r_+}, 
\end{equation}
where the effective Wilson coefficients are given by
\begin{eqnarray}
C_{\mathrm{eff},1}^F &=& C_1 +\frac{C2}{N}, \ C_{\mathrm{eff},2}^F =
C_2 +\frac{C_1}{N}, \ C_{\mathrm{eff},3}^F = C_3+\frac{C_4}{N},
\nonumber \\  
C_{\mathrm{eff},4}^F &=& C_4 +\frac{C_5}{N}, \ C_{\mathrm{eff},5}^F =
C_5 +\frac{C_6}{N}. 
\end{eqnarray}

The final expression can be simplied when we use the definition of
$f_+$ given in Eq.~(\ref{fplus}). For example, $T_1^p+ F_1^p$ is
written as
\begin{eqnarray}
T_1^p+ F_1^p &=&im_B^2 \zeta f_{\pi} \Bigl[\Bigl(C_1 +\frac{C_2}{N}\Bigr)
  (1+K) +\frac{\alpha_s}{4\pi} \frac{C_F}{N} C_2 F \Bigr], \nonumber
  \\ 
&&+i \frac{C_F}{N} \pi \alpha_s f_{\pi}^2 f_B m_B \Bigl(C_1
  +\frac{C_2}{N}\Bigr) \int 
du\frac{\phi_{\pi} (u)}{u} \int dr_+ \frac{\phi_B^+ (r_+)}{r_+}
  \nonumber \\
&=& im_B^2 \zeta f_{\pi} \frac{\alpha_s}{4\pi} \frac{C_F}{N} C_2 F  +
  im_B^2 f_{\pi} \Bigl( C_1 +\frac{C_2}{N} \Bigr) \Bigl[ \zeta (1+K) 
  \nonumber \\
&&  +\pi  \alpha_s \frac{C_F}{N} \frac{f_{\pi} f_B}{m_B}  \int 
du\frac{\phi_{\pi} (u)}{u} \int dr_+ \frac{\phi_B^+ (r_+)}{r_+} \Bigr]
  \nonumber \\
&&\approx im_B^2 f_+ f_{\pi}  \Bigl[\Bigl(C_1 +\frac{C_2}{N}\Bigr)
 +\frac{\alpha_s}{4\pi} \frac{C_F}{N} C_2 F \Bigr],
\end{eqnarray}
where the definition of $f_+$ in Eq.~(\ref{fplus}) is used in the last
line with $m_b \approx m_B =2E$. And we replace $\zeta$ by $f_+$ in the term
proportional to $F$. This induces terms of $\mathcal{O}
(\alpha_s^2)$, which is neglected. This relation also
holds for the combinations  $T_i^p+ F_i^p$ for all $i$.

In summary, the decay amplitudes $a_i^p$ are given by
\begin{eqnarray}
a_1^p &=& im_B^2 f_+ f_{\pi} \Bigl[\Bigl(C_1 +\frac{C_2}{N}\Bigr)
  +\frac{\alpha_s}{4\pi} \frac{C_F}{N} C_2 F^{\prime} \Bigr], \nonumber
  \\ 
a_2^p &=&im_B^2 f_+ f_{\pi} \Bigl[
\Bigl(C_2 +\frac{C_1}{N}\Bigr)   
+\frac{\alpha_s}{4\pi} \frac{C_F}{N} C_1 F^{\prime} \Bigr],\nonumber \\
a_3^p &=&im_B^2 f_+ f_{\pi} \Bigl[ \Bigl(C_3 +\frac{C_4}{N}\Bigr)
   +\frac{\alpha_s}{4\pi} \frac{C_F}{N} C_4 F^{\prime} \Bigr], \nonumber
  \\ 
a_4^p  &=& im_B^2 f_+ f_{\pi} \Biggl\{ \Bigl[\Bigl(C_4
  +\frac{C_3}{N}\Bigr)   
+\frac{\alpha_s}{4\pi} \frac{C_F}{N} \Bigl[ C_3 F^{\prime}  
+ C_1 \Bigl( \frac{2}{3}  -G(s_p) \Bigr) \nonumber \\ 
&+&C_3\Bigl( \frac{4}{3}   -G(0)
  -G(1) \Bigr) + C_4 \Bigl( \frac{2n_f}{3}  -3G(0)
  -G(s_c) -G(1) \Bigr) \nonumber \\
&+& C_6 \Bigl( -3G(0)
  -G(s_c) -G(1) \Bigr) +G_{\pi,8} (C_5+C_8) \Bigr] \Biggr\}, \nonumber \\
a_5^p &=&-im_B^2 f_+ f_{\pi} \Bigl[
\Bigl(C_5 +\frac{C_6}{N}\Bigr)   
+\frac{\alpha_s}{4\pi} \frac{C_F}{N} C_6 (-F^{\prime}-12) \Bigr],
\label{final}
\end{eqnarray}
where
\begin{equation}
F^{\prime} = F +  \frac{4\pi^2}{N} \frac{f_{\pi} f_B}{f_+ m_B^2} m_B 
\int_0^1 dr_+ \frac{\phi_B^+ (r_+)}{r_+} \Bigl( \int_0^1
  \frac{\phi_{\pi} (u)}{u} \Bigr)^2.
\end{equation}
The decay amplitudes in Eq.~(\ref{final}) at order $\alpha_s$ are
consistent with the result obtained by Beneke et al. \cite{bbns} at
order $\alpha_s$. However, our result goes further in the sense
that we consider the operators to all orders in $\alpha_s$, as shown
in Eqs.~(\ref{gen1}), (\ref{gen2}) and (\ref{gen3}). At higer
orders, our result will be different from the result in
Ref.~\cite{bbns} since there are two scales $m_b$ and $\mu_0=\sqrt{m_b
\Lambda}$ involved and the effects of these two scales cannot be
reproduced in the heavy quark limit.

\section{Conclusion\label{sec8}}
We have considered the four-quark operators relevant to nonleptonic $B$
decays into two light mesons in SCET at leading order in
$\Lambda$ and to next-to-leading order in $\alpha_s$. The construction
of the four-quark operators in SCET is process-dependent since we
first have to specify the directions of the outgoing quarks and
construct the operators accordingly. In matching onto
$\mathrm{SCET}_{\mathrm{II}}$, we integrate out off-shell
modes by attaching collinear and soft gluons to fermion lines. The
result is given as gauge-invariant four-quark operators. The form of
the gauge-invariant operators is obtained to all orders in $\alpha_s$
by using the auxiliary field method. The Wilson
coefficients of these operators can be computed by matching the
amplitudes between the full theory and $\mathrm{SCET}_{\mathrm{I}}$
since the infrared divergence of the full theory is reproduced in
$\mathrm{SCET}_{\mathrm{I}}$. And we obtain jet functions through the
matching between $\mathrm{SCET}_{\mathrm{I}}$ and
$\mathrm{SCET}_{\mathrm{II}}$.

When the effects of collinear and 
soft gluons are included, we can obtain gauge-invariant operators, and
the explicit form of these operators guarantees the color transparency
at leading order in $\Lambda$ but to all orders in $\alpha_s$. Now the
idea of the naive factorization in which the matrix elements of
four-quark operators are reduced to products of the matrix elements of
two currents has a theoretical basis.  Furthermore when we include
spectator interactions which contribute to the nonfactorizable
contribution and to the heavy-to-light form factor, the amplitudes,
which include four-quark and six-quark operators, are
factorized to all orders in $\alpha_s$. That is, the amplitudes can
be written as a convolution of short-distance effects represented by
the effective Wilson coefficients and long-distance effects
represented by the light-cone wave functions of mesons. And the
convolution integrals appearing in the hard spectator interactions and in
the hard contribution to the form factors are finite. Therefore we
have proved that the decay amplitudes for nonleptonic $B$ decays into two
light mesons at leading order in SCET and to all orders in $\alpha_s$
are factorized.   

Note that we have not included renormalization group running of the
effective Wilson coefficients. In order to include the renormalization
group running, the scaling of the Wilson coefficients
can be achieved by calculating the anomalous dimensions of the
operators, say, from Eqs.~(\ref{mat1}) and (\ref{mat5}). But the
running of the Wilson coefficients may not be appreciable at this
order for the scale change from $\mu=m_b$ to $\mu=\sqrt{m_b \Lambda}$.

As an application of SCET, we have calculated the decay amplitudes
for $\overline{B} \rightarrow \pi \pi$ at leading order in SCET. The
results are consistent at order $\alpha_s$ with the  result
in the heavy quark mass limit, which is shown explicitly here.
This is not a coincidence because the leading-order decay
amplitudes in the heavy quark mass limit employing leading-twist meson
wave functions correspond to the leading-order decay amplitudes in
SCET. However SCET extends the analysis to all orders in 
$\alpha_s$, and proves that the decay amplitudes are factorized. The
two types of the factorization properties, which 
correspond to the color transparency and the separation of long- and
short-distance effects are satisfied to all orders in $\alpha_s$ in
nonleptonic $B$ decays.

We can go beyond the leading-order calculation and consider subleading
 corrections  in order to check the validity of the approach
using SCET. For example, we can ask questions on how chirally-enhaced
 contributions can be treated in SCET, or how to include higher-twist
 wave functions of mesons, and how we can organize higher-order
 corrections in SCET.  However  we stress that this is a first step
 toward understanding nonleptonic $B$ decays, and those questions are
 under investigation.

\begin{ack}
This work was supported by Korea Research Foundation Grant
(KRF-2002-041-C00052). 
\end{ack}

\begin{appendix}
\section{Derivation of the gauge-invariant four-quark operators using
  the auxiliary field method} 
In Section~\ref{sec2} we derived gauge-invariant four-quark operators
through the matching calculation to order $g^2$, in which all the
off-shell modes were integrated out. We can derive the gauge-invariant
four-quark operators of the form $(\overline{\xi} \Gamma_1 h)\cdot
(\overline{\chi} \Gamma_2 \chi)$ using the auxiliary field method. 
We construct a Lagrangian of the on-shell fields with the off-shell
fields which are auxiliary fields. Then we integrate out the off-shell
auxiliary fields to obtain gauge-invariant operators entirely in terms
of the on-shell fields. In integrating out the off-shell modes, the
Wilson lines $W$, $\overline{W}$, $S$ and $\overline{S}$ appear and
the final form is gauge invariant. The auxiliary field method
offers a consistent way to derive gauge-invariant operators to all
orders.

The auxiliary field method has been used in
SCET for the gauge-invariant heavy-to-light current operators
\cite{bauer6}, and for the processes in which there are collinear
particles in the $n^{\mu}$ and $\overline{n}^{\mu}$ directions
\cite{bauer4}. We extend the idea of the auxiliary field method to
derive the gauge-invariant four-quark operators in nonleptonic
decays. Since there are collinear fields both in the $n^{\mu}$ and the
$\overline{n}^{\mu}$ directions, the auxiliary field method for
collinear quarks $\xi$ and $\chi$ is the same as the method presented
in Ref.~\cite{bauer4}. Here we consider how to treat the  heavy quark
interacting with two types of collinear gluons $A_n^{\mu}$ and
$A_{\overline{n}}^{\mu}$. 

For clarity, we first construct the Lagrangian of the on-shell fields
with the off-shell fields of order $p^2 \sim Q^2$, and integrate out
the off-shell modes to obtain the gauge-invariant four-quark
operators. Through this procedure, we obtain the Lagrangian and the
four-quark operators in $\mathrm{SCET}_{\mathrm{I}}$. Once we get
familiar with the construction of the Lagrangian with the off-shell
modes of order $p^2 \sim Q^2$, we include the off-shell modes of order
$p^2 \sim Q\Lambda$, and integrate out all the off-shell modes. We can
obtain the Lagrangian and the four-quark operators in
$\mathrm{SCET}_{\mathrm{II}}$ in a gauge-invariant way. 

Let us categorize all the fields in $\mathrm{SCET}_{\mathrm{I}}$
in powers of $\Lambda$. Note that the small expansion 
parameter $\lambda$ in $\mathrm{SCET}_{\mathrm{I}}$ is of order
$\sqrt{\Lambda/Q}$ and $\lambda^{\prime}$ in
$\mathrm{SCET}_{\mathrm{II}}$ is of order $\Lambda/Q$. In order to avoid
confusion, we express the scaling of all the momenta in powers of
$\Lambda$. The on-shell fields are the collinear fields $\xi$,
$A_n^{\mu}$, which scale as $p_n^{\mu} = (n\cdot p, \overline{n}\cdot
p,p_{\perp}) \sim (\Lambda, Q, \sqrt{Q\Lambda})$, the collinear fields
$\chi$, $A_{\overline{n}}^{\mu}$, which scale as $( Q, \Lambda,
\sqrt{Q\Lambda})$, the soft fields $q_s$, $A_s^{\mu}$, which scale as
$p_s^{\mu} \sim \sqrt{Q\Lambda}$, and the usoft fields  
$q_{us}$, $A_{us}^{\mu}$,  $h$, which scale as $p_{us}^{\mu}\sim
\Lambda$.

In $\mathrm{SCET}_{\mathrm{I}}$, we include the following off-shell
fields. When the on-shell heavy quark $h$ interacts with
a collinear gluon $A_n^{\mu}$ ($A_{\overline{n}}^{\mu}$), the final 
heavy quark has momentum of order $p_n^{\mu} \sim (\Lambda, Q,
\sqrt{Q\Lambda})$ [$p_{\overline{n}}^{\mu}\sim ( Q, \Lambda,
\sqrt{Q\Lambda})$]. We label this off-shell heavy quark as $\psi_n$
($\psi_{\overline{n}}$). If the on-shell heavy quark $h$ interacts
with $A_n^{\mu}$ and $A_{\overline{n}}^{\mu}$, the final off-shell
heavy quark has momentum of order $p_n^{\mu} + p_{\overline{n}}^{\mu}
\sim (Q,Q,\sqrt{Q\Lambda})$, and $p^2\sim Q^2$. We label this
off-shell field as $\psi_H$. A gluon
interacting with $A_n^{\mu}$ and $A_{\overline{n}}^{\mu}$ through the
triple gluon vertex also has a momentum $p_n^{\mu} +
p_{\overline{n}}^{\mu}$, and we label this gluon as $A_Q^{\mu}$. 
We list all the on-shell fields and the off-shell auxiliary fields in 
Table~\ref{intmod}. The collinear quarks can also be off-the mass
shell of order $p^2 \sim Q^2$, which we denote as $\xi_Q$ and
$\chi_Q$. The treatment of the off-shell collinear quarks is presented
in Ref.~\cite{bauer4}, and we will not repeat it here.
Since the soft and usoft quarks and gluons are not
relevant in our argument, we do not present them in Table~\ref{intmod}.

\begin{table}[t]
\caption{\label{intmod} List of fields to construct the auxiliary
  Lagrangian in $\mathrm{SCET}_{\mathrm{I}}$. The on-shell fields are
  collinear, and heavy quark fields, and the Wilson lines obtained by
  the corresponding gluons 
  are listed in the last column. The off-shell fields, which appear as
  intermediate states, are classified by their momentum scaling
  behavior.  } 
\vspace{0.3cm}
\begin{tabular}{cccc} \hline
& Momentum scaling& Fields& Wilson lines \\  \hline
on-shell& $p_n^{\mu}\sim (\Lambda, Q, (Q\Lambda )^{1/2})$
& $\xi$, $A_n^{\mu}$& $W$ \\ 
& $p_{\overline{n}}^{\mu} \sim
(Q,\Lambda,(Q\Lambda)^{1/2})$& $\chi$, 
$A_{\overline{n}}^{\mu}$ & $\overline{W}$ \\
& $p_{us}^{\mu} \sim (\Lambda, \Lambda,\Lambda)$& 
 $h$ &  \\ \hline 
off-shell&$p_n^{\mu} +p_{\overline{n}}^{\mu}\sim (Q,Q,
(Q\Lambda)^{1/2})$&   $\psi_H$, $A_Q^{\mu}$, $\xi_Q$, $\chi_Q$& $W_Q$,
$\overline{W}_Q$ 
\\  
& $p_n^{\mu}\sim (\Lambda, Q, (Q\Lambda )^{1/2})$&
$\psi_n$& \\
& $p_{\overline{n}}^{\mu}\sim (Q,\Lambda,(Q\Lambda)^{1/2})$
&$\psi_{\overline{n}}$& \\ 
\hline 
\end{tabular}
\end{table}

The auxiliary gluon Lagrangian in $\mathrm{SCET}_{\mathrm{I}}$ is
given by 
\begin{equation}
\mathcal{L}^g_{\mathrm{aux}} [A_Q] = \frac{1}{2g^2} \mathrm{tr}
\Bigl( [iD_Q^{\mu} +gA_Q^{\mu}, iD_Q^{\nu} +gA_Q^{\nu}]\Bigr)^2
+\frac{1}{\alpha_L} \mathrm{tr} \Bigl(\Bigl[iD_{Q\mu}, A_Q^{\mu}]
\Bigr)^2,
\label{gaux}
\end{equation}
where the covariant derivative $D_Q^{\mu}$ is given by
\begin{eqnarray}
iD_Q^{\mu} &=&\frac{n^{\mu}}{2}(\overline{n} \cdot \mathcal{P} +
g \overline{n} \cdot A_n) +\frac{\overline{n}^{\mu}}{2}
(n\cdot \mathcal{Q} +gn\cdot A_{\overline{n}})
\nonumber \\
&&+\mathcal{P}_{\perp}^{\mu} +gA_{n\perp}^{\mu}
+\mathcal{Q}_{\perp}^{\mu} +g A_{\overline{n}\perp}^{\mu},
\label{cov}
\end{eqnarray}
and only the leading terms in Eq.~(\ref{cov}) is
included in $\mathcal{L}_{\mathrm{aux}}^g$ at leading order. 

We can separately obtain the solution of the first term and the second
term in Eq.~(\ref{gaux}). The equation of motion of
the first term is given by
\begin{equation}
[iD_{Q\mu} +gA_{Q\mu} ,[iD_Q^{\mu} + gA_Q^{\mu}, iD_Q^{\nu}
    +gA_Q^{\nu}]] =0.
\label{eom}
\end{equation}
The leading-order solution is obtained by making an ansatz 
\begin{equation}
\overline{W}_Q^{\dagger} W_Q = W \overline{W}^{\dagger}.
\label{lan}
\end{equation}
Here $W_Q$ and $\overline{W}_Q$ are essentially the Fourier transform
of the Wilson lines
\begin{eqnarray}
W_Q (y) &=& \mathrm{P} \exp \Biggl\{ ig \int_{-\infty}^y ds \Bigl[
  \overline{n} \cdot A_Q (s\overline{n}) +\overline{n} \cdot A_n
  (s\overline{n}) \Bigr] \Biggr\}, \nonumber \\
\overline{W}_Q (y) &=& \mathrm{P} \exp \Biggl\{ ig \int_{-\infty}^y ds
  \Bigl[ n \cdot A_Q (sn) +n \cdot A_{\overline{n}} (sn) \Bigr]
  \Biggr\}, 
\label{gluong}
\end{eqnarray}
which satisfy
\begin{equation}
\overline{n} \cdot \Bigl(\mathcal{P} +gA_Q+g A_n \Bigr) W_Q =0, \
n \cdot \Bigl(\mathcal{Q} +g A_Q +g A_{\overline{n}}
\Bigr) \overline{W}_Q =0.
\end{equation}

The Lagrangian with auxiliary heavy fields is complicated because only
the sum of three graphs shown in Eq.~(\ref{heavy}) is simple, but the
individual diagrams show complex behavior. Recall that
Eq.~(\ref{heavy}) is given by
\begin{equation}
M_a + M_b =\frac{ig^2}{2} f_{abc} \overline{q} \Gamma_1 T_a \frac{n\cdot
  A_{\overline{n}}^b \overline{n} \cdot A_n^c}{n\cdot q_{\overline{n}}
  \overline{n} \cdot q_n}b+ g^2 \overline{q} \Gamma_1
  \frac{\overline{n} \cdot A_n}{\overline{n} 
  \cdot q_n} \frac{n\cdot A_{\overline{n}}}{n\cdot q_{\overline{n}}}
  b,
\label{heavy1}
\end{equation}
and we construct the Lagrangian
such that the first term in Eq.~(\ref{heavy1}) is reproduced by the
auxiliary field attached with the triple gluon vertex with $A_n^{\mu}$
and $A_{\overline{n}}^{\mu}$, and the second term is produced by
attaching $A_n^{\mu}$ and $A_{\overline{n}}^{\mu}$ starting from the
vertex in this order. It means that when a heavy quark $h$ becomes an
off-shell field $\psi_H$ interacting with $A_n^{\mu}$ and
$A_{\overline{n}}^{\mu}$, the heavy field $h$ first interacts with
$A_{\overline{n}}^{\mu}$ to become $\psi_{\overline{n}}$, and
then interacts with $A_n^{\mu}$ to become $\psi_H$. We can choose a
different convention in which $h$ interacts with $A_n^{\mu}$ first and
then interacts with $A_{\overline{n}}^{\mu}$ to become $\psi_H$. Both
methods give the same result, as will be shown later. However, it is
necessary to specify the order of the interaction 
since the Lagrangian for the off-shell fields is obtained by expanding
the QCD Feynman rules in powers of $\Lambda$.

With the prescription described above, the Lagrangian for the heavy
quark with the auxiliary fields is given as 
\begin{eqnarray}
\mathcal{L}_{\mathrm{aux}}^h  &=& \overline{\psi}_n g\overline{n}\cdot
  A_n h   +\overline{\psi}_{\overline{n}} gn\cdot A_{\overline{n}} h
  +\overline{\psi}_H g\overline{n}   \cdot A_Q  (h+\psi_n) \nonumber
  \\ 
&&+\overline{\psi}_H
  g\overline{n} \cdot (A_Q + A_n )
  \psi_{\overline{n}}  + \mathrm{h.c.}
  \nonumber \\
&&+\overline{\psi}_n (\overline{n}\cdot \mathcal{P} +g\overline{n}
  \cdot A_n)   \psi_n + \overline{\psi}_{\overline{n}} (n\cdot
  \mathcal{Q} +gn \cdot A_{\overline{n}} )\psi_{\overline{n}}
  \nonumber   \\
&&+\overline{\psi}_H \Bigl[\overline{n} \cdot \mathcal{P} +g\overline{n}
  \cdot (A_Q + A_n ) \Bigr] \psi_H. 
\label{hqaux}
\end{eqnarray}
Note that there is no term in which $\psi_n$ interacts with
$A_{\overline{n}}^{\mu}$ to become $\psi_H$ according to our
prescription. The inclusion of the auxiliary field $A_Q^{\mu}$ in the
second line of Eq.~(\ref{hqaux}) should be verified at order $g^3$,
but these terms are included since they are kinematically allowed.

Solving for $\psi_{\overline{n}}$, $\psi_n$ and $\psi_H$ from
Eq.~(\ref{hqaux}),  we obtain 
\begin{eqnarray}
&&gn\cdot A_{\overline{n}} h +n\cdot (\mathcal{Q} + gA_{\overline{n}} )
\psi_{\overline{n}} =0, \label{eqmo} \\
&&g\overline{n}\cdot A_n h +\overline{n} \cdot (\mathcal{P}
+g A_n )\psi_n =0, \nonumber \\
&& g\overline{n} \cdot A_Q (h +\psi_n) +g\overline{n}\cdot (A_n +A_Q)
\psi_{\overline{n}}  + \overline{n} \cdot \Bigl[\mathcal{P} +
g (A_n +A_Q) \Bigr] \psi_H =0. \nonumber
\end{eqnarray}
The first equation in Eq.~(\ref{eqmo}) can be solved for
$\psi_{\overline{n}}$ as 
\begin{equation}
\psi_{\overline{n}} = (\overline{W} -1)h,
\end{equation}
and adding the second and the third equations in Eq.~(\ref{eqmo})
yields
\begin{equation}
\psi_n +\psi_H = (W_Q-1) (h+\psi_{\overline{n}}).
\end{equation}
Therefore the heavy quark field can be written as
\begin{equation}
h+\psi_{\overline{n}}+\psi_n +\psi_H = h+\psi_{\overline{n}} + (W_Q-1)
(h+\psi_{\overline{n}}) = W_Q \overline{W} h.
\label{finh}
\end{equation}

We can arrive at an equivalent conclusion by rewriting
Eq.~(\ref{heavy1}) as
\begin{equation}
M_a + M_b =\frac{ig^2}{2} f_{abc} \overline{q} \Gamma_1 T_a \frac{n\cdot
  A_{\overline{n}}^c \overline{n} \cdot A_n^b}{n\cdot q_{\overline{n}}
  \overline{n} \cdot q_n}b+ g^2 \overline{q} \Gamma_1
  \frac{n\cdot A_{\overline{n}}}{n\cdot q_{\overline{n}}}
  \frac{\overline{n} \cdot A_n}{\overline{n} b 
  \cdot q_n},
\label{heavy2}
\end{equation}
where the order of the gluons in the second term is reversed. In this
case, we require that the heavy quark $h$ interacts with $A_n^{\mu}$
first and then interacts with $A_{\overline{n}}^{\mu}$ to generate
$\psi_H$. The corresponding Lagrangian is given by
\begin{eqnarray}
\mathcal{L}_{\mathrm{aux}}^h  &=& \overline{\psi}_n g\overline{n}\cdot
  A_n h   +\overline{\psi}_{\overline{n}} gn\cdot A_{\overline{n}} h
  +\overline{\psi}_H gn   \cdot A_Q  (h
  +\psi_{\overline{n}})  \nonumber \\
&&+\overline{\psi}_H
  gn \cdot (A_Q + A_{\overline{n}})  \psi_n +\mathrm{h.c.} 
  \nonumber \\
&&+\overline{\psi}_n (\overline{n}\cdot \mathcal{P} +g\overline{n}
  \cdot A_n)   \psi_n + \overline{\psi}_{\overline{n}} (n\cdot
  \mathcal{Q} +gn \cdot A_{\overline{n}} )\psi_{\overline{n}}
  \nonumber   \\
&&+\overline{\psi}_H \Bigl[n \cdot \mathcal{Q} +gn
  \cdot (A_Q + A_{\overline{n}} ) \Bigr] \psi_H. 
\label{hqaux1}
\end{eqnarray}
Solving the equations of motion, we obtain 
\begin{equation}
h+\psi_n+\psi_{\overline{n}}  +\psi_H = \overline{W}_Q W h = W_Q
\overline{W} h,
\end{equation}
where the last equality comes from the ansatz 
$\overline{W}_Q^{\dagger} W_Q = W\overline{W}^{\dagger}$. This is
equivalent to the result in Eq.~(\ref{finh}).

\begin{table}[b]
\caption{\label{modscet} List of fields to construct the auxiliary
  Lagrangian in $\mathrm{SCET}_{\mathrm{II}}$. The fields
  with the index $X$ respresent the off-shell fields 
  with the soft momentum. The momentum scaling of $\psi_H$,
  $A_Q^{\mu}$, $\psi_n$ and $\psi_{\overline{n}}$  does not change
  with the addition of the soft momentum.} 
\vspace{0.3cm}
\begin{tabular}{cccc} \hline
& Momentum scaling& Fields& Wilson lines \\  \hline
on-shell& $p_n^{\mu}\sim (\Lambda^2/Q, Q, \Lambda)$
& $\xi$, $A_n^{\mu}$& $W$ \\ 
& $p_{\overline{n}}^{\mu} \sim
(Q,\Lambda^2/Q,\Lambda)$& $\chi$, 
$A_{\overline{n}}^{\mu}$ & $\overline{W}$ \\
& $p_{s}^{\mu} \sim (\Lambda, \Lambda,\Lambda)$& $q_{s}$,
$A_{s}^{\mu}$, $h$ & $S$, $\overline{S}$\\ \hline 
off-shell&$p_n^{\mu} +p_{\overline{n}}^{\mu}+p_s\sim (Q,Q,
\Lambda)$&   $\psi_H$, $A_Q^{\mu}$& $W_Q^X$,   $\overline{W}_Q^X$ \\  
&$p_n^{\mu} +p_s^{\mu}\sim (\Lambda, Q, \Lambda)$ &
$\psi_n$, $\xi_X$, $A_{nX}^{\mu}$ & $W_X$, $S_X$ \\ 
&$p_{\overline{n}}^{\mu} +p_s^{\mu} \sim
(Q,\Lambda,\Lambda)$ & $\psi_{\overline{n}}$,
$\chi_X$,  $A_{\overline{n} X}^{\mu}$ & $\overline{W}_X$,
$\overline{S}_X$ \\ 
\hline 
\end{tabular}
\end{table}

Now we construct the Lagrangian including the off-shell modes of order
$p^2 \sim Q\Lambda$. Through this procedure, we obtain the Lagrangian
and the four-quark operators in $\mathrm{SCET}_{\mathrm{II}}$. 
In $\mathrm{SCET}_{\mathrm{II}}$, we have to integrate out all the
off-shell modes in Table~\ref{modscet} by constructing the Lagrangian with
additional auxiliary fields with additional soft momentum, with the
index $X$. The on-shell and the off-shell fields in
$\mathrm{SCET}_{\mathrm{II}}$ are listed in Table~\ref{modscet}. Note
that the momentum scaling of each fields has changed in order to
accomodate $\mathrm{SCET}_{\mathrm{II}}$. Since $p_n^{\mu} +
p_{\overline{n}}^{\mu} + p_s^{\mu} \sim p_n^{\mu} +
p_{\overline{n}}^{\mu}$, we do not put the index $X$ to $\psi_H$ and
$A_Q^{\mu}$. For the same reason, we express the off-shell heavy
quarks as $\psi_n$ and $\psi_{\overline{n}}$.  

We include the off-shell modes due to the soft momentum. We
label these fields with the index $X$ to denote the off-shellness 
$p^2\sim Q\Lambda$. The auxiliary Lagrangian including the
off-shell modes of order $p^2 \sim Q\Lambda$ is given by
\begin{eqnarray}
\mathcal{L}_{\mathrm{aux}}^h  &=& \overline{\psi}_n g\overline{n}
  \cdot (A_n^X + A_n) h   +\overline{\psi}_{\overline{n}} gn\cdot
  (A_{\overline{n}}^X +A_{\overline{n}}) h
  +\overline{\psi}_H g\overline{n}   \cdot A_Q  (h+\psi_n) \nonumber 
  \\ 
&&+\overline{\psi}_H
  g\overline{n} \cdot (A_Q + A_n^X+ A_n )
  \psi_{\overline{n}}  + \mathrm{h.c.}
  \nonumber \\
&&+\overline{\psi}_n \overline{n}\cdot  \Bigl[\mathcal{P} +g
 (A_n^X+ A_n) \Bigr]   \psi_n 
+ \overline{\psi}_{\overline{n}} n\cdot\Bigl[
  \mathcal{Q} +g (A_{\overline{n}}^X +A_{\overline{n}} ) \Bigr]
  \psi_{\overline{n}}
  \nonumber   \\
&&+\overline{\psi}_H \overline{n} \cdot \Bigl[\mathcal{P} +g (A_Q +
  A_n ) \Bigr] \psi_H.  
\label{hqQ}
\end{eqnarray}
Note that there is no soft gluon $A_s^{\mu}$ involved in
Eq.~(\ref{hqQ}) for the heavy quark. It appears in the Lagrangian for
the collinear quarks. The explicit Lagrangian for the collinear quarks
involving the off-shell fields of order $p^2 \sim Q\Lambda$ can be
found in Eqs.~(A4) and (A5) in Ref.~\cite{bauer4}.

The difference between the Lagrangians given in Eqs.~(\ref{hqaux}) and
(\ref{hqQ}) is that all the off-shell fields are labeled by $X$, and
$A_n\rightarrow A_n^X + A_n$, $A_{\overline{n}} \rightarrow
A_{\overline{n}}^X + A_{\overline{n}}$. Therefore the heavy quark
field, after the off-shell fields are integrated out, is written as
\begin{equation}
h +\psi_n + \psi_{\overline{n}} +\psi_H = W_Q^X \overline{W}_X h,  
\label{trans}
\end{equation}
and the collinear quarks are written as \cite{bauer4}
\begin{equation}
\xi +\xi_X + \xi_Q = \overline{W}_Q^X
S_X \xi, \ \chi +\chi_X + \chi_Q = W_Q^X \overline{S}_X \chi,
\end{equation}
where $W_Q^X$, and $S_X$ are the Fourier transforms of the Wilson lines
\begin{eqnarray}
W_Q^X (y) &=& \mathrm{P} \exp \Biggl\{ ig \int_{-\infty}^y ds \Bigl[
  \overline{n} \cdot A_Q (s\overline{n}) + \overline{n} \cdot A_{nX}
  (s\overline{n}) +
\overline{n} \cdot A_n
  (s\overline{n}) \Bigr] \Biggr\}, \nonumber \\
S_X (y) &=& \mathrm{P} \exp \Biggl\{ ig \int_{-\infty}^y ds
  \Bigl[ n \cdot A_{nX} (sn) +n \cdot A_s (sn) \Bigr]
  \Biggr\},
\label{gluong2}
\end{eqnarray}
and $\overline{W}_Q^X$ and $\overline{S}_X$ are obtained by replacing
$n^{\mu}$ by $\overline{n}^{\mu}$ in Eq.~(\ref{gluong2}).
The Wilson lines satisfy
\begin{equation}
\overline{W}_Q^{X\dagger} W_Q^X = W_X \overline{W}_X^{\dagger}, \
S_X^{\dagger} W_X = WS^{\dagger}, \ \overline{S}_X \overline{W}_X =
\overline{W} \overline{S}^{\dagger}.
\label{wilson}
\end{equation}
The last two relations in Eq.~(\ref{trans}) were obtained in
Ref.~\cite{bauer4} and the first relation is new.

By integrating out the off-shell fields, the singlet four-quark
operator in $\mathrm{SCET}_{\mathrm{II}}$ is given by
\begin{eqnarray}
&& \Bigl(\overline{\xi} +\overline{\xi}_X +\overline{\xi}_Q \Bigr)
  \Gamma_1   \Bigl(h+\psi_n 
  +\psi_{\overline{n}} +\psi_H \Bigr)  \nonumber \\
&&\times \Bigl(\overline{\chi}
  +\overline{\chi}_X +\overline{\chi}_Q \Bigr) \Gamma_2 \Bigl(\chi
  +\chi_X   +\chi_Q \Bigr)  \nonumber \\
&=& \Bigl(\overline{\xi} S_X^{\dagger} \overline{W}_Q^{X\dagger}
  \Gamma_1 W_Q^X   \overline{W}_X h \Bigr) \cdot \Bigl(
  \overline{\chi} \overline{S}_X^{\dagger}  W_Q^{X\dagger} \Gamma_2
  W_Q^X \overline{S}_X \chi \Bigr)\nonumber \\ 
&=& \Bigl(\overline{\xi} S_X^{\dagger} W_X \Gamma_1
  \overline{W}_X^{\dagger} \overline{W}_X h \Bigr) \cdot \Bigl( \chi
  \Gamma_2 \chi\Bigr) =\overline{\xi} W \Gamma_1 
  S^{\dagger} h \cdot \overline{\chi}   \Gamma_2 \chi,
\end{eqnarray}
using Eq.~(\ref{wilson}), and similarly the nonsinglet four-quark
operator is given as 
\begin{eqnarray}
&& \Bigl(\overline{\xi} S_X^{\dagger} \overline{W}_Q^{X\dagger}
  \Bigr)_{\beta} \Gamma_1 \Bigl( W_Q^X \overline{W}_X h\Bigr)_{\alpha}
  \cdot \Bigl( \overline{\chi} \overline{S}_X^{\dagger} W_Q^{X\dagger}
  \Bigr)_{\alpha}   \Gamma_2 \Bigl(W_Q^X \overline{S}_X
  \chi\Bigr)_{\beta} \nonumber \\
&& = \Bigl(\overline{\xi} S_X^{\dagger} W_X \Bigr)_{\beta} \Gamma_1
  \Bigl( \overline{S}_X^{\dagger} \overline{W}_X h\Bigr)_{\alpha}
  \cdot \overline{\chi}_{\alpha} \Gamma_2 \Bigl(
  \overline{W}_X^{\dagger} \overline{S}_X \chi\Bigr)_{\beta} \nonumber
  \\
&&= ( \overline{\xi} WS^{\dagger} )_{\beta} \Gamma_1 (
  \overline{S}^{\dagger} h)_{\alpha} \cdot (\overline{\chi}
  \overline{W} )_{\alpha} \Gamma_2 (\overline{S}
  \overline{W}^{\dagger} \chi)_{\beta} \nonumber \\
&&= \Bigl(\overline{\xi} WS^{\dagger} \overline{S} \Bigr)_{\beta}
  \Gamma_1 ( \overline{S}^{\dagger} h )_{\alpha} \cdot
  ( \overline{\chi} \overline{W} )_{\alpha} \Gamma_2 (
  \overline{W}^{\dagger} \chi)_{\beta}.
\end{eqnarray}
This result is the same as the explicit calculation obtained in
Section 2, and it is a proof to all orders in $\alpha_s$ using the
auxiliary field method.

\section{Derivation of the subleading operators using the auxiliary
  field method}
We can also derive the subleading operators $O^{(1a,1b)}_i$ in
Eq.~(\ref{o1ab}) and $J^{(1a)}_i$ and $J^{(1b)}_i$ in
Eq.~(\ref{fsubop}) using the auxiliary field method. Here we have to
consider off-shell modes from the collinear gluons both in the
$n^{\mu}$ and $\overline{n}^{\mu}$ directions. For simplicity,
we consider the subleading operators in
$\mathrm{SCET}_{\mathrm{I}}$ disregarding the off-shell modes by soft
gluons. The leading-order result was derived in
Ref.~\cite{bauer4}, and here we present a new result which yields
gauge-invariant operators at subleading order.  In addition to the
solutions in Eq.~(\ref{eom}) at leading order, when we include the
subleading terms in the covariant derivative in Eq.~(\ref{cov}), we
obtain the equations of motion at subleading order, which are given as   
\begin{eqnarray}
[W_Q \overline{n} \cdot \mathcal{P} W_Q^{\dagger}, [\overline{W}_Q
    n\cdot \mathcal{Q} \overline{W}_Q^{\dagger},
    i\mathcal{D}_{Q\perp}^{\nu} ]]     &=&0, \nonumber \\ 
{}[\overline{W}_Q n\cdot \mathcal{Q}
    \overline{W}_Q^{\dagger}, [W_Q 
\overline{n} \cdot \mathcal{P} W_Q^{\dagger},
    i\mathcal{D}_{Q\perp}^{\nu} ]]     &=&0,
\label{subeom}
\end{eqnarray}
where $i\mathcal{D}_{Q\perp}^{\mu} = iD_{Q\perp}^{\mu}
+gA_{Q\perp}^{\mu}$.  
Here we make an ansatz 
\begin{equation}
\overline{W}_Q^{\dagger} i\mathcal{D}_{n\perp Q}^{\nu} =
iD_{n\perp}^{\nu} 
\overline{W}_Q^{\dagger} \ \mbox{or} \ W_Q^{\dagger}
i\mathcal{D}_{\overline{n}\perp Q}^{\nu} = iD_{\overline{n}\perp}^{\nu}
W_Q^{\dagger}, 
\label{san}
\end{equation}
and they satisfy Eq.~(\ref{subeom}).

First the intermediate form of the subleading four-quark operators
relevant to the nonfactorizable spectator contribution and the
heavy-to-light form factor can be obtained by neglecting the off-shell
modes in the $\overline{n}^{\mu}$ direction since we are
interested in the operators proportional to $gA_n^{\mu}$ at leading
order in $g$. As a definite example, let us concentrate on the 
operator $O^{(1b)}_i$ and obtain the gauge-invariant form by integrating
out the off-shell modes. The derivation for the operator $O^{(1a)}_i$
can be done in a similar way. In 
$\mathrm{SCET}_{\mathrm{I}}$, without the off-shell modes in the
$\overline{n}^{\mu}$ direction, the operator $O^{(1b)}_i$ is given by
\begin{eqnarray}
O^{(1b)}_i &=& (\overline{\xi} +\overline{\xi}_Q) \Gamma_{1i}  (h
+\psi_n  +\psi_{\overline{n}} + \psi_H) \nonumber \\
&&\times (\overline{\chi} +\overline{\chi}_Q) \Gamma_{2i}
\Bigl( \overline{W}_Q \frac{1}{n\cdot \mathcal{Q}}
  \overline{W}_Q^{\dagger}\Bigr)   [
 i\FMSlash{D}_{n\perp} +g\FMSlash{A}_{Q\perp} ] \frac{\FMslash{n}}{2}
(\chi +\chi_Q), 
\label{sample}
\end{eqnarray}
where the color indices are suppressed. When we include off-shell
modes, each field is written as
\begin{equation}
h +\psi_n + \psi_{\overline{n}} = W_Q \overline{W} h, \ \xi +\xi_Q=
\overline{W}_Q \xi, \ \chi +\chi_Q  = W_Q \chi.
\end{equation}
Then the operator in Eq.~(\ref{sample}) is written as
\begin{equation} 
\Bigl[ (\overline{\xi} \overline{W}_Q^{\dagger}) \Gamma_{1i} (W_Q
  \overline{W} h) \Bigr]
\cdot \Bigl[(\overline{\chi} W_Q^{\dagger}) \Gamma_{2i} 
\overline{W}_Q \frac{1}{n\cdot \mathcal{Q}} \overline{W}_Q^{\dagger}
  \Bigl( [i\FMSlash{\mathcal{D}}_{n\perp Q} 
  W_Q] \frac{\FMslash{n}}{2} \chi \Bigr) \Bigr].  
\end{equation}

If we use the ansatz in Eqs.~(\ref{lan}) and (\ref{san}), the singlet
operator becomes 
\begin{eqnarray}
O_S^{(1b)} &=& \Bigl[ (\overline{\xi} \overline{W}_Q^{\dagger}
  W_Q)_{\alpha}   \Gamma_1 (\overline{W} h)_{\alpha} \Bigr] \nonumber
  \\ 
&&\times \Bigl[(\overline{\chi} W_Q^{\dagger})_{\beta} \Gamma_2 
\Bigl( \overline{W}_Q \frac{1}{n\cdot \mathcal{Q}}
  \overline{W}_Q^{\dagger} [i\FMSlash{\mathcal{D}}_{n\perp Q} W_Q]
\frac{\FMslash{n}}{2} \chi \Bigr)_{\beta} \Bigr] \nonumber \\
 &=& \Bigl[ (\overline{\xi} W)_{\alpha}
  \Gamma_1 h_{\alpha} \Bigr] \nonumber \\
&&\times \Bigl[(\overline{\chi} W_Q^{\dagger}\overline{W}_Q)_{\beta}
  \Gamma_2  
\frac{1}{n\cdot \mathcal{Q}} \Bigl([i\FMSlash{D}_{n\perp}
  \overline{W}_Q^{\dagger} W_Q]
\frac{\FMslash{n}}{2} \chi \Bigr)_{\beta} \Bigr] \nonumber  \\
&=&\Bigl[(\overline{\xi} W)_{\alpha} \Gamma_1 h_{\alpha} \Bigr] \cdot 
\Bigl[( \overline{\chi} 
\overline{W})_{\beta}  \Gamma_2 \frac{1}{n\cdot \mathcal{Q}}\Bigl(
  [W^{\dagger}   i\FMSlash{D}_{n\perp} W] \frac{\FMslash{n}}{2}
\overline{W}^{\dagger} \chi \Bigr)_{\beta} \Bigr], 
\end{eqnarray}
and the nonsinglet operator becomes
\begin{eqnarray}
O_N^{(1b)} &=& \Bigl[ (\overline{\xi}
  \overline{W}_Q^{\dagger})_{\beta}   \Gamma_1 (W_Q \overline{W}
  h)_{\alpha} \Bigr] \cdot 
\Bigl[(\overline{\chi} W_Q^{\dagger})_{\alpha} \Gamma_2 \Bigl(
  \overline{W}_Q \frac{1}{n\cdot \mathcal{Q}} \overline{W}_Q^{\dagger}
  [i\FMSlash{\mathcal{D}}_{n\perp Q} W_Q] 
\frac{\FMslash{n}}{2} \chi \Bigr)_{\beta} \Bigr] \nonumber \\
&=&\Bigl[ \overline{\xi}_{\beta}
  \Gamma_1 (\overline{W} h)_{\alpha} \Bigr] \cdot
\Bigl[\overline{\chi}_{\alpha} \Gamma_2 
\frac{1}{n\cdot \mathcal{Q}} \Bigl( \overline{W}_Q^{\dagger}
  [i\FMSlash{\mathcal{D}}_{n\perp Q} W_Q]  
\frac{\FMslash{n}}{2} \chi \Bigr)_{\beta} \Bigr] \nonumber \\ 
&=& \Bigl[ \overline{\xi}_{\beta}
  \Gamma_1 (\overline{W} h)_{\alpha} \Bigr] \cdot
\Bigl[\overline{\chi}_{\alpha} \Gamma_2 
\frac{1}{n\cdot \mathcal{Q}} \Bigl( 
  [i\FMSlash{D}_{n\perp} \overline{W}_Q^{\dagger} W_Q]  
\frac{\FMslash{n}}{2} \chi \Bigr)_{\beta} \Bigr] \nonumber \\ 
&=& \Bigl[( \overline{\xi} W\Bigr)_{\beta} \Gamma_1
h_{\alpha}\Bigr] \cdot \Bigl[ \Bigl(\overline{\chi} \overline{W}
  \Bigr)_{\alpha} \Gamma_2 \Bigl([W^{\dagger} i\FMSlash{D}_{n\perp}
  W] \frac{1}{n \cdot   \mathcal{Q}} \frac{\FMslash{n}}{2}
\overline{W}^{\dagger} \chi\Bigr)_{\beta}\Bigr],
\end{eqnarray}
These are the subleading operators which are gauge invariant.
Other operators can be shown to have the forms presented in the paper
in a similar way. 
\end{appendix}

\end{document}